# Analog to digital conversion in beam instrumentation systems

*M. Gasior*
Beam Instrumentation Group, CERN, Geneva, Switzerland

**Abstract**
Analog to digital conversion is a very important part of almost all beam instrumentation systems. Ideally, in a properly designed system, the used analog-to-digital converter (ADC) should not limit the system performance. However, despite recent improvements in ADC technology, quite often this is not possible and the choice of the ADC influences significantly or even restricts the system performance. It is therefore very important to estimate the requirements for the analog to digital conversion at an early stage of the system design and evaluate whether one can find an adequate ADC fulfilling the system specification. In case of beam instrumentation systems requiring both, high time and amplitude resolution, it often happens that the system specification cannot be met with the available ADCs without applying special processing to the analog signals prior to their digitisation. In such cases the requirements for the ADC even influence the system architecture. This paper aims at helping the designer of a beam instrumentation system in the process of selecting an ADC, which in many cases is iterative, requiring a trade-off between system performance, complexity and cost. Analog to digital conversion is widely and well described in the literature, therefore this paper focusses mostly on aspects related to beam instrumentation. The ADC fundamentals are limited to the content presented as an introduction during the CAS one-hour lecture corresponding to this paper.

## 1 Introduction

Analog to digital conversion is a vast subject described in thick textbooks and lectured over semesters, therefore selecting topics for one-hour lecture was a difficult task. It could not be assumed that the audience knew ADCs basics, so by necessity the lecture had to start with a short introduction. Then the next presented topics were selected to at least inform the potential designer of a beam instrumentation system which matters should be taken into account during the design process. Preference was given to the subjects likely to be encountered during an early stage of a system design, which are less popular in literature and more related to fundamental aspects than the current state of the technology. This paper follows the lecture strategy. After this brief introduction, ADC basics are presented, followed by a discussion on ADC fundamental limitations. Then the very important subject of signal sampling is presented along with even more important deliberations on which sampling rate is needed in a beam instrumentation system, followed by the corresponding discussion on the required ADC resolution. The next two chapters treat the more practical aspects of choosing a good ADC for a beam instrumentation system and options for obtaining ADC boards. The paper finishes with a summary and a short list of literature.

For general aspects related to ADCs that were not possible to accommodate in this paper, like their architectures, the Reader is referred to textbooks. An excellent and comprehensive handbook on data conversion can be downloaded [1]. Textbooks [2, 3], published in many languages, are very good for learning about data converters as well as electronics in general. It is recommended to download and read a paper [4] and the corresponding two-hour lecture [5] on ADCs and DACs from the 2008 Beam Diagnostics CAS in Dourdan, France. To some extent this paper and the related CAS lecture were designed to complement the excellent content presented by Jeroen Belleman during the 2008 CAS.

Unfortunately, the one-hour BI CAS lecture could not include digital to analog conversion and digital-to-analog converters (DACs), as this is a vast subject on its own, and the paper follows the same strategy. This choice is also justified by the fact that DACs are not often critical parts of beam instrumentation systems. On the other hand, some ADC types, like successive approximation ADCs, use digital to analog conversion in the analog to digital conversion process. This is why a Reader interested in deeper understanding ADCs is invited to also study DACs, which are well described in textbooks [1-3] and the mentioned CAS material [4, 5].

## 2   ADC basics

The principle of analog to digital conversion can be illustrated with a simple example of a perfect, 4-bit ADC with a parallel output bus, a 5 V reference voltage and a 1 MHz sampling clock; the ADC symbol is shown in Fig. 1(a). Let's imagine that the ADC input signal is a triangular waveform shown in Fig. 1(b). The first fundamental operation performed by the ADC is sampling, that is taking amplitudes of the input signal at the instants defined by the sampling clock, while the signal between the samples is ignored. This is why sampling has fundamental consequences: the input signal can be faithfully reconstructed from its samples only when certain conditions mentioned later are fulfilled.

The second ADC operation is quantisation, that is each sample amplitude is expressed by an integer number. Our example ADC has 4 bits allowing representing $2^4 = 16$ values, from 0 (binary states '0000' on the output bus) to 15 (states '1111'). The reference voltage is 5 V, so one digital unit (quantum) at the ADC output, the least significant bit (LSB), corresponds to $V_{LSB}$ = 5 V / 16 = 0.3125 V. The digitisation process consists of assigning to each sample amplitude an integer number corresponding to the voltage being its nearest multiple of $V_{LSB}$. Here the consequences are much simpler than in the case of sampling: the quantisation process introduces a systematic error, often called quantisation error, being the difference between the analog input amplitudes and their corresponding integer multiplies of $V_{LSB}$. Please note that the worst-case quantisation error in a perfect ADC is just half of $V_{LSB}$.

The quantisation process of the samples of our triangle waveform is illustrated in Fig. 1(c). It can be seen in Fig. 1(d) that the input signal maximum is larger than the ADC input dynamic range of 5 V, saturating the ADC and resulting in an additional error, larger than half of $V_{LSB}$. Once the signal is sampled and quantised, it can be represented by a sequence of integer numbers, as show in Fig. 1(e), and the integer numbers are represented by binary signals on the output bus, as shown in Fig. 1(f). The clock signal rising edges indicate when the waveform amplitudes are sampled and the corresponding integers registered by the following digital system that is receiving and processing the ADC data. Please note that this example, for the sake of simplicity, ignores many details, for example the delay introduced by the ADC circuitry between the analog input signal and the output digital data.

A perfect 4-bit ADC has conversion characteristic as presented in Fig. 2(a), showing the staircase function according to which the input analog signal is converted into the output digital data. Characteristics of real converters always have errors and some of their examples are shown in Fig. 2. These are examples of "static" ADC errors, while ADCs also have dynamic imperfections, some of which are mentioned later. Depending the ADC application, certain errors may be more important than others. For example, in systems with ADCs measuring DC signals the offset error is an important parameter, while in systems optimised for spectral analysis this parameter is in most cases not at all important. In general, which errors one should expect depends on the architecture of the ADC and this is the domain where knowledge of different ADC architectures may help. However, ADC errors are specified in their datasheets and as such should be sufficient to quantify the ADC performance, regardless of its architecture. For recent, high-performance ADCs their architectures are very complex and often combine classical "textbook" architectures to gain in performance. This is another reason why the ADC architectures are not treated in this paper.

Now let's introduce the most important ADC parameters. If a perfect converter with *n* bits has symmetric, bipolar dynamic range ± *A*, then the analog amplitude corresponding to the change on the digital output by one unit is

$$V_{LSB} = \frac{2A}{2^n} \quad (1)$$

As illustrated on the example in Fig. 3, the largest quantisation error is

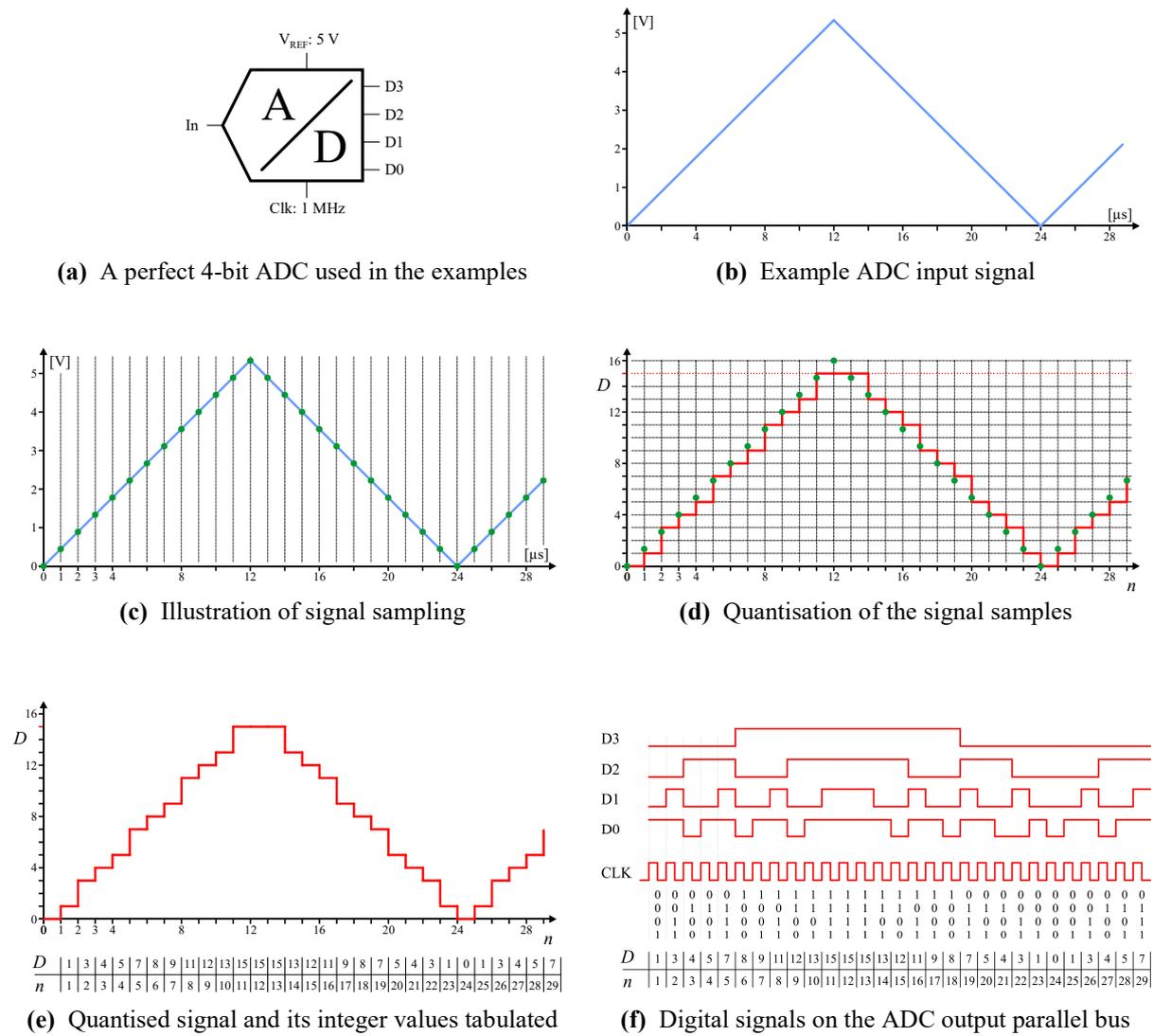

**Fig. 1:** A perfect 4-bit ADC used in examples (a), along with signals illustrating the operations involved in analog to digital conversion (b) – (f).

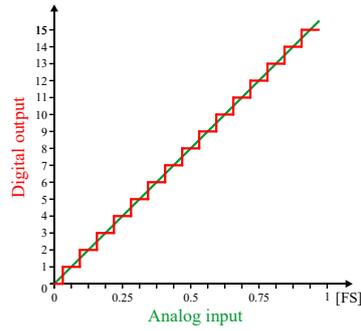

**(a)** Ideal characteristic of a 4-bit ADC

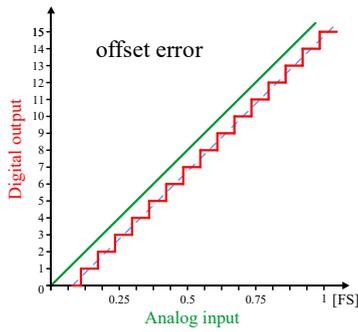

**(b)** An illustration of an offset error

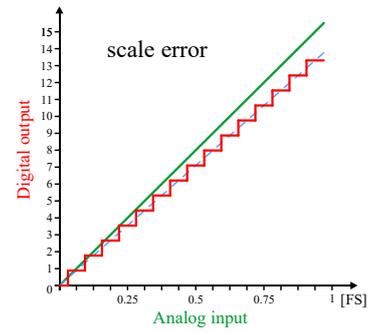

**(c)** An illustration of a scale error

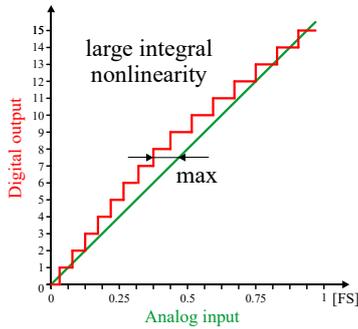

**(d)** An illustration of an integral nonlinearity

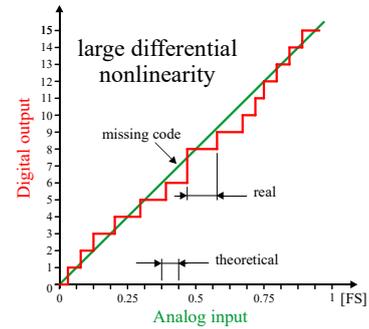

**(e)** An illustration of a differential nonlinearity

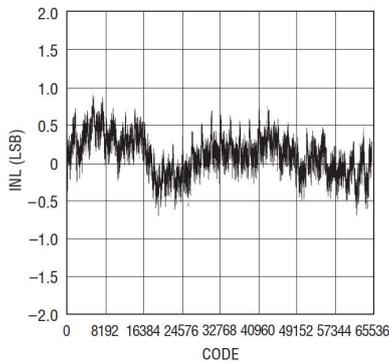

**(f)** An example of datasheet integral nonlinearity

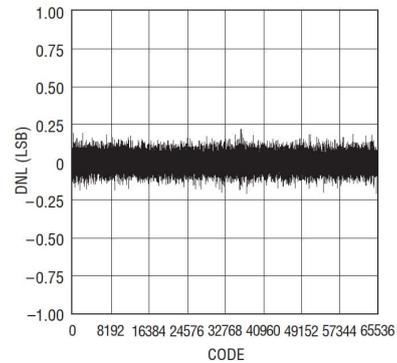

**(g)** An example of datasheet differential nonlinearity

**Fig. 2:** Ideal characteristic of a 4-bit ADC (a) and illustrations of common ADC static errors (b) – (e), along with datasheet integral nonlinearity (f) and differential nonlinearity (g) for the LTC2205.

$$e_m = \frac{V_{LSB}}{2} \qquad (2)$$

The root mean square (RMS) of the quantisation error is

$$e_{rms} = \frac{V_{LSB}}{\sqrt{12}} \cong 0.29\, V_{LSB} \qquad (3)$$

and the RMS amplitude of a sinusoidal signal with amplitude $A$ is

$$A_{rms} = \frac{A}{\sqrt{2}} \qquad (4)$$

Using two quantities $e_{rms}$ (3) and $A_{rms}$ (4) one can define many of the ADC parameters that often appear in ADC datasheets [6].

Signal-to-noise ratio (SNR) is defined as the quotient of the signal and noise RMS amplitudes. Substituting (1), (2) and (4) one gets:

$$SNR = \frac{A_{rms}}{e_{rms}} = \frac{\sqrt{6}}{2} 2^n \qquad (5)$$

Often SNR is given in decibels (dB) and then it is a linear function of the number of bits $n$:

$$SNR\,[dB] = 20 \log_{10} \frac{\sqrt{6}}{2} 2^n \cong 1.76 + 6.02\, n \qquad (6)$$

The equation (6) above is used "backwards" to define another important ADC parameter. Imagine that we have a real ADC with measured signal-to-noise ratio $SNR_M$. We can then calculate the so-called effective number of bits (ENOB) of this ADC. This quantity says how many bits a perfect ADC, having the same SNR as our real one, would have:

$$ENOB = \frac{SNR_M\,[dB] - 1.76}{6.02} \qquad (7)$$

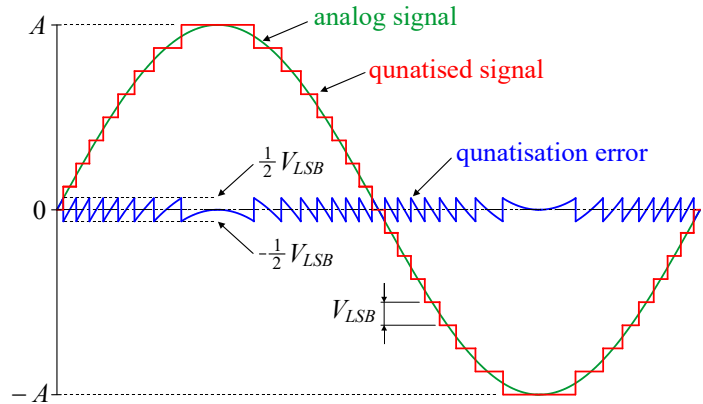

**Fig. 3:** An illustration of the quantisation error, which is the difference between the original analog signal and its quantised equivalent. $A$ and $V_{LSB}$ are the amplitudes of the analog signal and the quantisation step, respectively.

Sometimes ENOB can be defined using not the measured SNR but the measured signal-to-noise and distortion parameter $SINAD_M$, and then:

$$ENOB' = \frac{SINAD_M \,[dB] - 1.76}{6.02} \tag{8}$$

SINAD takes into account not only the noise of the ADC, but also spurious components introduced solely by the ADC due to its distortion of the sinusoidal input signal. However, this definition of ENOB leads to smaller numbers in datasheets and therefore is not often used by ADC manufacturers.

When the ADC input signal is a well-defined shape, the quantisation error looks also like a waveform, as in Fig. 3. In general, for real input signals, the quantisation error looks like noise and is then often called quantisation noise. Figure 4 shows a numerical simulation of a perfect 4-bit ADC. Its simulated input signal is 10 000 full-scale random numbers (green dots), which are then converted into integer numbers using the round() mathematical function (red dots). The quantisation error is the difference between the "analog" real numbers and "digital" integer numbers (blue dots). The calculated RMS amplitude of the quantisation error for this example is 0.287, while the theoretical value calculated from (3) is 0.289.

Employing sinusoidal input signals and analysing ADC data in the frequency domain reveals many details of ADCs performance, which would be difficult to notice otherwise. This approach is used to quantify many ADC parameters presented in datasheets as well as to characterise real systems with ADCs. The strength of this approach is illustrated in the following numerical examples, still related to our perfect 4-bit ADC.

In the first example one simulates a full-scale sine wave with frequency $f_h$ of 1 % of the sampling frequency $f_s$ to use this signal as the ADC input. One period of this signal and the corresponding ADC output are shown in Fig. 5(a). The magnitude of Fourier spectrum of $N$ = 10 000 ADC samples is shown in Fig. 5(b). The plot has a logarithmic vertical scale in dB, normalised to the fundamental component. The expected spectral component corresponding to the input sinusoidal signal is seen at the discrete spectrum bin of index

$$k_{in} = N \frac{f_{in}}{f_s} = 100 \tag{9}$$

However, there are also almost all odd harmonics of the input signal present in the spectrum. The highest one has a level of – 32 dB (41st harmonic), resulting in a 32 dB spurious-free dynamic range (SFDR) for the case of our perfect 4-bit ADC. Please note that for a different relationship between the input signal and the sampling instants the result could be different. Curious Readers are encouraged to extend presented numerical simulations and to also check other cases.

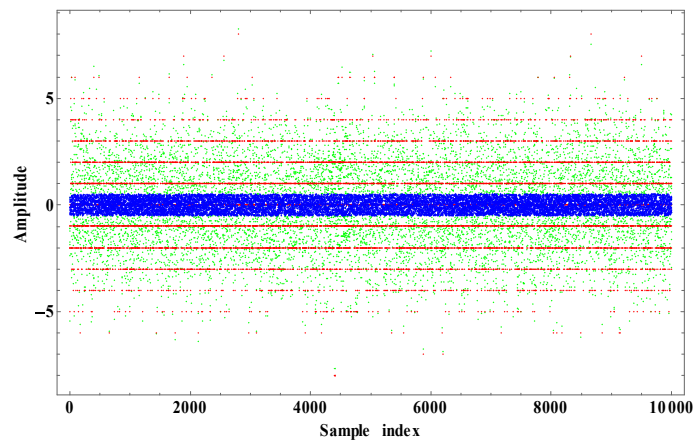

**Fig. 4:** Results of a numerical simulation of a perfect 4-bit ADC with random input values represented by green dots. ADC output data is marked as red dots and the quantisation error as blue dots.

The observed spurious components are present due to the fact that the digitisation process is perfect and the quantisation error is a waveform, which results in the observed spectrum. For the next example shown in Fig. 5(c) the same sine waveform (black) was used as the ADC input signal, but now with an addition of random numbers with a Gaussian distribution and standard deviation of about

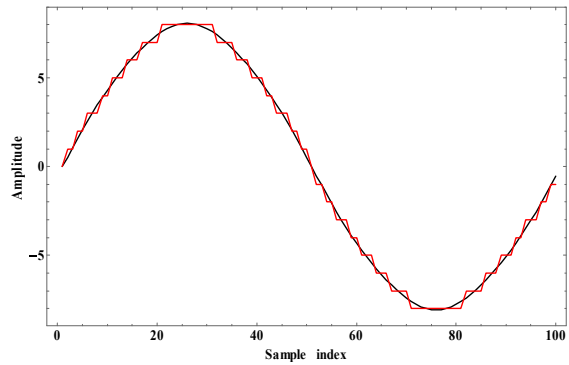

**(a)** One period of the signal and its 4-bit samples

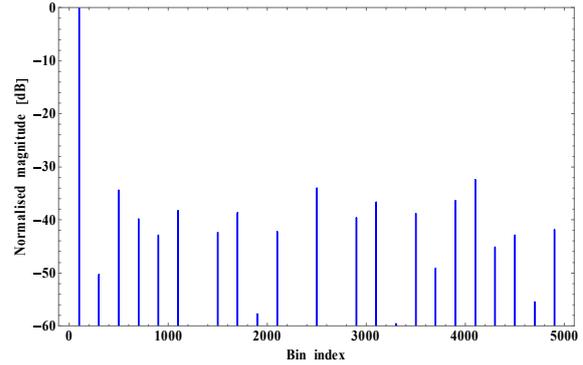

**(b)** Magnitude spectrum of the 4-bit samples (a)

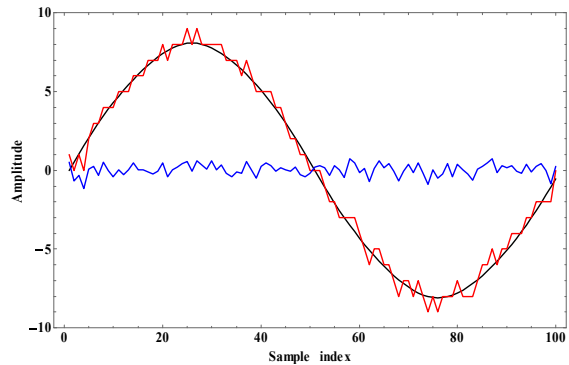

**(c)** Analysed signal with additive noise

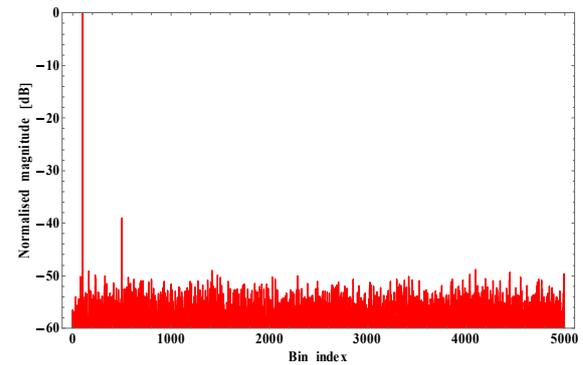

**(d)** Magnitude spectrum of 4-bit samples (c)

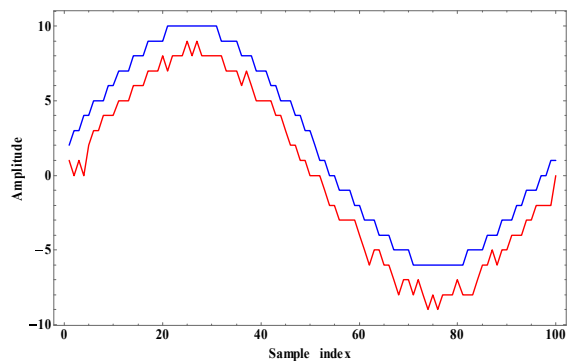

**(e)** Comparison of samples (a) and (c)

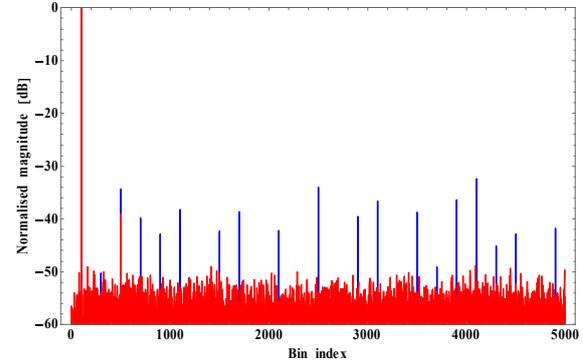

**(f)** Comparison of spectra (b) and (d)

**Fig. 5:** Examples of quantisation error spectra. 4-bit samples (a) do not contain any noise, so the corresponding spectrum (b) contains spurious components. When noise is added (c), then the spurious spectral lines disappear (d). Plots (e) and (f) show the comparisons of the two cases without and with the additional noise.

2.5 $V_{LSB}$ (blue), simulating an additive input noise. In this case the spectrum shown in Fig. 5(d) does not have obvious spurious components, except one, but rather a noise floor.

The unexpected component in bin 500 in fact corresponds to a second small sinusoidal component that was silently added to the input signal. Its frequency is exactly 5 times the frequency of the fundamental tone, so in the spectrum in Fig. 5(b) it is combined with the 5$^{th}$ spurious harmonic, and therefore it could be smuggled to the spectrum without being noticed. Now with the added noise this component is nicely visible in spite of the fact that its amplitude is only 1 % of the fundamental tone. Please note that its spectral level is 40 dB lower than the fundamental, corresponding to the expected factor 100. With the mystery of the second component explained, one can conclude that the noise addition increased the SFDR by 17 dB (a factor of 7), from the previous 32 dB to 49 dB, as the largest spectral component has now the level of – 49 dB.

Comparisons of the two cases, without and with the added noise, are shown in Fig. 5(e) and (f). In fact, it is remarkable that adding quite a bit of noise improves performance of such an ADC when it is used in a system employing spectral analysis. As the spurious components are suppressed, more dynamic range is made available for detecting small spectral content. In this case, the quantisation noise of larger energy is spread more or less evenly over all bins, instead of being condensed in several ones, increasing the ADC spurious-free dynamic range.

The examples presented also show the strength of spectral analysis as a tool for detecting small signal components. Please note that our 4-bit ADC, which can distinguish only 16 amplitudes, allowed detecting with a reasonable signal-to-noise ratio a component with the amplitude of 1 % of the ADC dynamic range, much below one LSB.

The shown numerical examples illustrate a technique called dithering, in which a small amount of noise is added to the ADC input signal to "randomise" the conversion process and to decrease spurious spectral components. Figure 6 shows the datasheet spectra of a real 16-bit ADC, without (left) and with dither (right). For this converter dithering is an option selected by putting a logic state on one of the chip inputs. As seen in Fig. 6, for this ADC dithering increases the SFDR by some 10 dB.

The examples presented are meant to demonstrate dithering, but their purpose is also to encourage the use of simple numerical simulations to study specific behaviour of ADCs in beam instrumentation systems. Please note that ADC data samples are just integer numbers and, as such, can be easily generated and processed using your favourite math application.

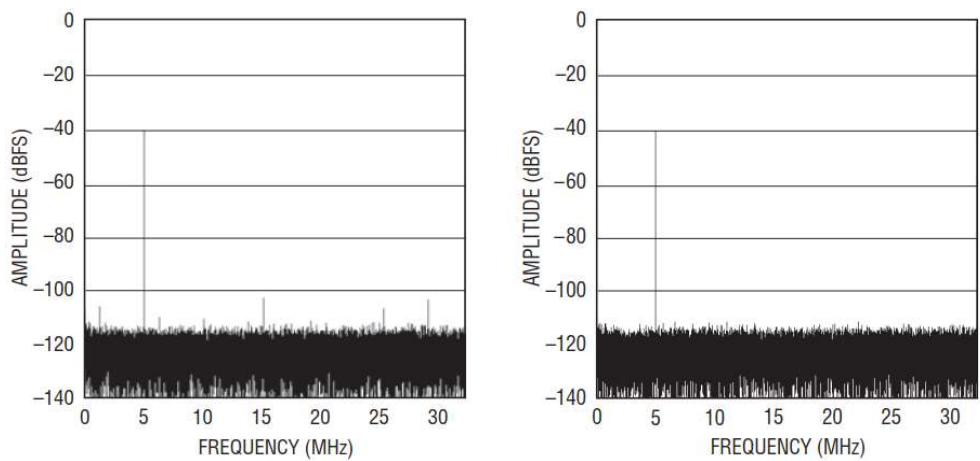

**Fig. 6:** Datasheet spectra without dithering (left) and with dithering (right) for LTC2205 sampling at 70 MHz. Spectra with 64 K points for 5.1 MHz input signal with 1 % amplitude of the ADC full scale.

Discrete spectra are often used to analyse ADC performance and it is important to know how the noise in the time and frequency domains is related. In the following paragraphs some math is used to explain this relationship and give examples how one can deal with time and frequency data, with the hope that it could be also an inspiration for calculating other quantities.

Assume that we have a signal containing a sinusoidal component with RMS amplitude of $V_c$ and frequency $f_c$. The signal also contains white noise with the RMS amplitude of $V_n$. Then the time domain SNR of the sinusoidal component is

$$SNR_t = \frac{V_c}{V_n} \tag{10}$$

If the signal is sampled at frequency $f_s$, its $N$ samples are acquired during time $NT_s$, where $T_s$ is the sampling period and $T_s = f_s^{-1}$, and the signal and noise respective energies are

$$E_c = V_c^2 NT_s \tag{11a}$$
$$E_n = V_n^2 NT_s \tag{11b}$$

The discrete spectrum of the signal, here means the magnitude of its discrete Fourier transform (DFT), has $N$ bins, among which $N/2$ carry the whole signal spectrum, while the other $N/2$ are just their symmetric copies. The width of one spectral bin is

$$\Delta_f = \frac{f_s}{N} \tag{12}$$

Now let's assume that the sinusoidal component frequency $f_c$ is an integer multiple of $\Delta_f$

$$f_c = k\frac{f_s}{N} = k\Delta_f \tag{13}$$

where $k$ is an integer and $k \leq N/2$. In this case the discrete spectrum contains only one bin (of index $k$) corresponding to the sinusoidal component.

From the Parseval's theorem we know that the energies of the sinusoidal component are equal in both the time and frequency domains. The energy of the component in the frequency domain is in one bin only, so the component spectral density is

$$\frac{dV_c}{df} = \sqrt{\frac{V_c^2 NT_s}{\Delta_f}} = NT_s V_c \tag{14}$$

The noise present in the signal is white, so it is distributed evenly over $N/2$ spectral bins. Again, the noise energy is equal in both, time and frequency domains, so the noise spectral density is

$$\frac{dV_n}{df} = \sqrt{\frac{V_n^2 NT_s}{\frac{N}{2}\Delta_f}} = \sqrt{2N}T_s V_n \tag{15}$$

Therefore, the signal-to-noise ratio of the sinusoidal component in the frequency domain, meaning the ratio of the respective spectral densities, is

$$SNR_f = \frac{NT_s V_h}{\sqrt{2N}T_s V_n} = \sqrt{\frac{N}{2}}\frac{V_h}{V_n} = \sqrt{\frac{N}{2}} SNR_t \tag{16}$$

This is why one can often see a quantity called "FFT gain", relating the signal-to-noise ratios in the time and frequency domains

$$\text{FFT gain} = \frac{SNR_f}{SNR_t} = \sqrt{\frac{N}{2}} \qquad (17)$$

The name "FFT gain" is related to the fact that in practice the discrete Fourier transform used to obtain signal spectra is almost always calculated using the fast Fourier transform (FFT) numerical algorithm.

If we have a signal with a time-domain signal-to-noise ratio $SNR_t$ and we perform spectral analysis on $N$ and $kN$ number of samples, then the quotient of the corresponding signal-to-noise ratios in the frequency domain calculated using (16) is

$$\frac{SNR_f(kN)}{SNR_f(N)} = \frac{\sqrt{kN/2}\ SNR_t}{\sqrt{N/2}\ SNR_t} = \sqrt{k} \qquad (18)$$

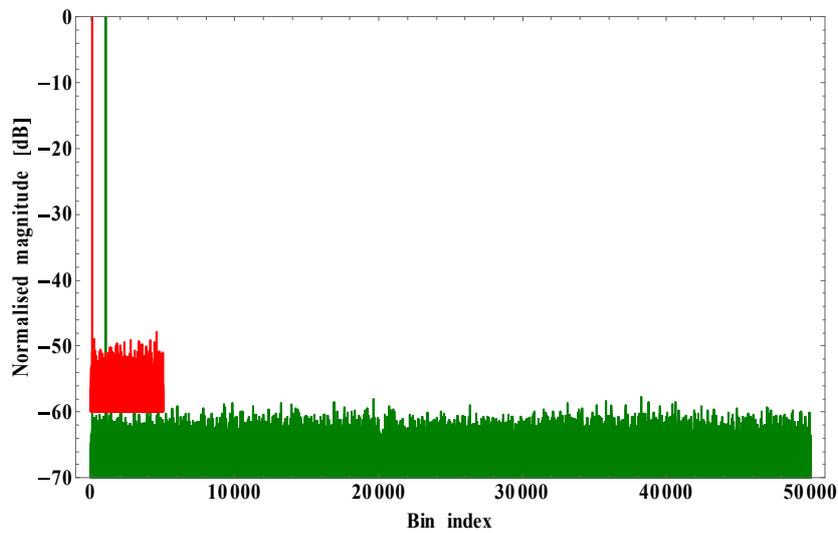

**Fig. 7:** Two spectra calculated upon 10 000 (red) and 100 000 (green) samples. The expected reduction of the noise floor for the longer spectrum is 10 dB and this is approximately the value seen.

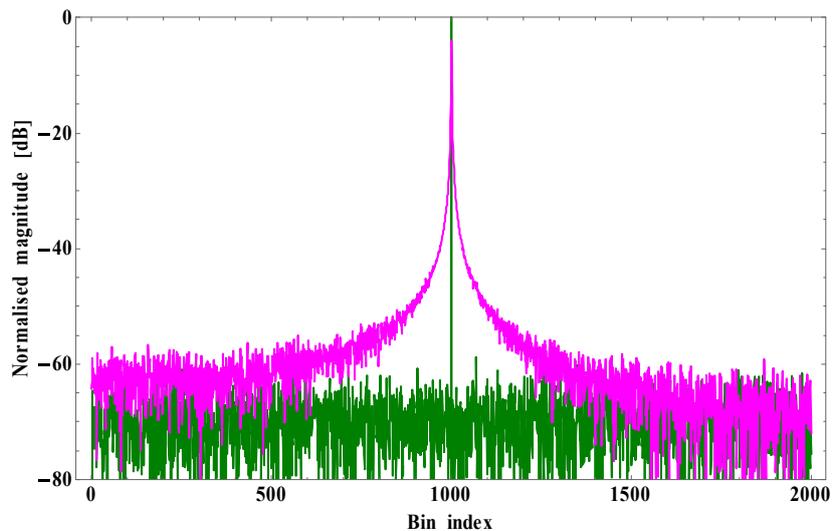

**Fig. 8:** A spectrum suffering from spectral leakage (magenta) and the corresponding spectrum without spectral leakage (green). Signal frequencies, normalised to the sampling rate, are 0.010005 and 0.01 respectively.

Thus, the signal-to-noise ratio in the frequency domain improves proportionally to the square root of the sample number increase. This is illustrated with a numerical example shown in Fig. 7 with two spectra. The first one (green) is very similar to the one used in the previous examples, calculated from $N$ = 10 000 4-bit samples of a full-scale sinusoidal component with a similar addition of noise. The sinusoidal component frequency $f_h$ is 1 % of the sampling frequency $f_s$. The second spectrum (red) corresponds to the same signal, but calculated from $N'$ = 100 000 = 10 $N$ samples. The number of samples increased by factor of 10, therefore, as calculated from (18), the signal-to-noise ratio is expected to improve by $20 \cdot \log_{10}(\sqrt{10})$ = 10 dB and this is approximately the value seen in the plot.

Please note that in the presented calculations it was assumed that the frequency of the analysed component is such that in the frequency domain it lies exactly on the discrete spectrum bin. If this is not the case, the component energy is distributed over many spectral bins, as illustrated with the spectrum shown in magenta in Fig. 8. The spectrum was calculated similarly to the previous example with 100 000 samples, which for comparison is shown in green, but the signal frequency was shifted by half of the bin spacing to maximise the spectral leakage. Please note that the peak of the spectrum with leakage is lowered, as the component energy in this case is spread over many bins.

Most ADC measurements, including those in datasheets, are performed with signals having frequencies with a relationship to the sampling rate in order to guarantee that no spectral leakage pollutes the measured spectra. There are ways to reduce the spectral leakage, for example signal windowing. However, with such techniques the improvement in the spectral leakage is traded for an increase of the width of the spectral peak related to the fundamental component. Consequently, there is a reduction of the signal-to-noise ratio and this is why such techniques are not welcome in measurements characterising ADC performance.

## 3 Fundamental ADC limitations

Progress in ADC technology and the resulting increase in ADC performance has been remarkable and this process is not likely to ever stop. However, there are phenomena that put asymptotic limits on many ADC parameters. In this chapter a few such limitations are described that are worth remembering during the design process of a beam instrumentation system. Some of the limitations are supported by numerical examples, meant to better reveal their origins.

### 3.1 Slew rate

Let's imagine that at the output of an amplifier there is a sinusoidal signal of the form

$$s = A \sin(2\pi f t) \qquad (19)$$

The signal maximal rate of change, called often the slew rate (SR), is

$$SR(s) = max\left(\frac{ds}{dt}\right) = 2\pi A f \qquad (20)$$

Therefore, as illustrated in Fig. 9, the slew rate of a signal depends on both, its amplitude and frequency.

The amplifier has its own slew rate limit, often specified in its datasheet. If the signal slew rate is faster than the amplifier slew rate limit, the amplifier is not able to follow and the signal will be distorted. The "distortion level" will depend on how much faster the signal is with respect to the amplifier limit. For an amplifier with a slew rate *SR* and a sinusoidal signal with amplitude $A_{max}$ the maximal frequency $f_{max}$ of the signal without distortion is therefore

$$f_{max} = \frac{SR}{2\pi A_{max}} \qquad (21)$$

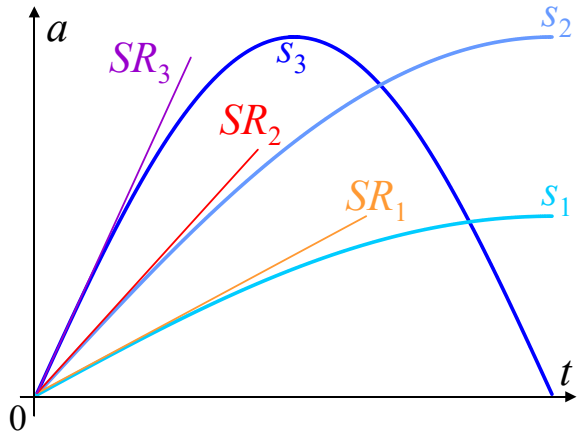

**Fig. 9:** Illustration of the slew rate concept. Slew rate depends on both, signal amplitude and signal frequency.

Fast ADCs require both the internal and external circuitry to have very large slew rates. Unfortunately, the slew rate is limited by the technology used and, as shown later, also by the affordable power dissipation. When the slew rate cannot be increased further and one still needs faster ADCs, then the only solution is to limit the maximal amplitude of the processed signals. This trend can be seen in ADC datasheets: while for slower ADCs the typical input dynamic range is 5 V, for the fastest ADCs the dynamic range can even be lower than 1 V. For some beam instrumentation systems this can limit the achievable signal-to-noise ratio. This will be the case if the system noise or interference level does not scale with the signal level at the ADC input and in real systems this is often true.

A common way of decreasing the required slew rate of the ADC input circuitry is to equip the ADC with differential inputs. Then each of the two differential input signals has half of the total amplitude seen by the ADC. The reduction of the required slew rate is achieved at the expense of more complex circuitry. In the extreme case the whole electronic chain before the ADC can be differential and built as two identical channels processing signals with the opposite polarities.

Differential ADC inputs are also beneficial for reducing interference and noise, which in many cases is quite similar on both sides of the differential input. The differential signals are subtracted inside the ADC and any common part of the two signals (the so-called "common mode") is reduced as seen by the ADC. The level to which the common part is supressed with respect to the differential signal is called common mode rejection ratio (CMRR) and is specified in ADC datasheets. Typically, the CMRR gets worse with increasing frequency of the input signals, as for faster signals it is more difficult to maintain the symmetry of the ADC differential paths.

In practice all high-performance ADCs have differential inputs. The differential architecture can either be kept for the whole processing chain in front of the ADC or the last stage before the ADC can provide the conversion from a simple "one-signal scheme" (a so-called single-ended architecture), to the differential input of the ADC.

At the time of writing this paper the fastest operational amplifiers available on the market have slew rates in the order of 10 V/ns. If such amplifiers are used to drive an ADC with a 5 V dynamic range and a differential input, then the maximal frequency of a sinusoidal signal without distortion calculated from (21) is about 640 MHz. Reducing the dynamic range to 1 V allows increasing this frequency to 3.2 GHz.

## 3.2 Power dissipation

The power dissipated by an ADC chip depends on its power supply voltage and the required operating current. The voltage sets the maximum ADC dynamic range. The larger the supply voltage, the larger dynamic range, but unfortunately, also the higher the power dissipation. Excessive power dissipation requires special means to evacuate the heat from the chip in order to keep its temperature within acceptable limits and assure reliable operation with the parameters declared in the datasheet. As a rule of thumb, one can assume that power dissipation in the order of 0.1 W and below, dissipated by a small ADC chip or any other component, is easy to handle, providing that there are not too many such components on the board, so that the total power dissipation is not excessive. Power dissipation around 1 W in an ADC chip requires some attention from the designer, especially for precision designs in which temperature is an important factor. As the power dissipation increases towards 10 W, dedicated studies are necessary on how to evacuate the corresponding heat from the circuit. Even larger power dissipation levels typically require active cooling techniques, for example, solutions similar to those used for computer CPU and GPU chips.

High-performance ADCs often have a few power supply voltages to limit the total power dissipation while maintaining a larger dynamic range. A typical case is three supply voltages:
- "analog voltage";
- "digital voltage";
- voltage for the input/output circuitry, here referred to as "I/O voltage".

The "analog voltage" (often 5 V or 3.3 V) is the highest and supplies only the input circuitry which is important for the dynamic range. This voltage is the most critical for the ADC performance, as its quality affects the noise and interference seen in the ADC data. ADC datasheets provide so-called power supply rejection ratio (PSRR), quantifying how much of the analog supply voltage noise or interference bleeds through to the output data. This parameter is similar to the CMRR mentioned earlier and is also a function of frequency.

The "digital voltage" supplies all of the digital circuits in the ADC and, as such, can be kept quite small, with typical values around 1 V. For high-performance ADCs the digital circuitry is often complex, requiring a lot of current, so keeping this voltage low reduces the total power dissipation.

The "I/O voltage" supplies the ADC digital input and output circuits, which are connected to the "master logic" that is taking data from the ADC and controlling its operation. The voltage is typically equal to the corresponding I/O voltage of the "master logic", which in many cases is a field-programmable gate array (FPGA). Typical I/O voltage values are between 3.3 V and 1.2 V, corresponding to standards for different logic chip families.

Optimising the supply current of an ADC is less straightforward. Let's assume that an internal signal in an ADC is a voltage $v_s$ and it is routed through a path with parasitic capacitance $C_p$. The rate of change of the voltage is

$$\frac{dv_s}{dt} = \frac{i_c}{C_p} \qquad (22)$$

where $i_c$ is the capacitance current. From this equation one can conclude that once the parasitic capacitances are minimised by optimising the chip production technology, the only way of increasing the speed of the circuitry is to increase the switching current. This is why very fast ADCs with GHz sampling rates have large supply currents and power dissipation.

To give an idea of the real-world numbers involved, let's assume that we need to drive a circuit with the parasitic input capacitance of 1 pF. We ask for the slew rate to be 10 V/ns, as for the fast operational amplifier from the previous "slew-rate example". From (22) we calculate that the needed charging current is 10 mA. A high-performance ADC is a complex system with many lines requiring fast drive, so the individual currents add up, resulting in a quite important total current required by the fast ADC.

The 1 pF capacitance assumed in this example is rather small. For example, even a good, fast oscilloscope can have on its inputs parasitic capacitances in the order of 10 pF.

## 3.3 Noise

The most fundamental origin of noise is that generated by resistances. A resistance $R$ with an absolute temperature $T$ in measurement bandwidth $B$ produces white noise with the RMS voltage

$$v_n = \sqrt{4kTRB} \tag{23}$$

where $k$ is the Boltzmann constant. Therefore, increasing the bandwidth of a system always means a larger system noise. Also, low noise circuits require small resistances to be present in the signal path, resulting in important currents, which consequently require a larger supply power. This is why it is difficult to make low noise electronics with low supply power and why micropower circuits are not likely to have very good noise performance.

Now let's have a look at some numbers corresponding to an extreme case, assuming that we have a noiseless ADC and the only noise source is a 50 Ω resistor at room temperature of 300 K. The resistor is on the ADC input and terminates a coaxial cable delivering the input signal. From (23) we can calculate the RMS noise voltage for a given measurement bandwidth, which here is considered as the bandwidth of a beam instrumentation system. A few example values are listed in Table 1 for bandwidths increasing from 1 Hz to 1 GHz in steps of three orders of magnitude, so for each step the noise voltage increases by factor $\sqrt{1000} \cong 31.6$ that is by $20 \cdot \log_{10}(1000^{0.5}) = 30$ dB. The next two columns list the corresponding signal-to-noise rations of a sinusoidal signal for two cases of peak-peak amplitudes of 5 V and 1 V, imagined to be the full-scale voltages of our hypothetical ADC. The last two columns list the effective number of bits for the listed SNRs.

**Table 1:** Theoretical ADC noise performance for different system bandwidths assuming that it is defined by an input 50 Ω resistor in temperature 300 K.

| Bandwidth | $V_n$ | $SNR_{5V}$ [dB] | $SNR_{1V}$ [dB] | $ENOB_{5V}$ [bit] | $ENOB_{1V}$ [bit] |
|---|---|---|---|---|---|
| 1 Hz | 0.9 nV | 186 | 172 | 30.6 | 28.2 |
| 1 kHz | 29 nV | 156 | 142 | 25.6 | 23.3 |
| 1 MHz | 910 nV | 126 | 112 | 20.6 | 18.3 |
| 1 GHz | 29 μV | 96 | 82 | 15.6 | 13.3 |

**Table 2:** Theoretical ADC noise performance for different system bandwidths assuming that it is defined by an input 5 kΩ resistor in temperature 300 K.

| Bandwidth | $V_n$ | $SNR_{5V}$ [dB] | $SNR_{1V}$ [dB] | $ENOB_{5V}$ [bit] | $ENOB_{1V}$ [bit] |
|---|---|---|---|---|---|
| 1 Hz | 9.1 nV | 166 | 152 | 27.2 | 24.9 |
| 1 kHz | 288 nV | 136 | 122 | 22.3 | 19.9 |
| 1 MHz | 9.1 μV | 106 | 92 | 17.3 | 15.0 |
| 1 GHz | 288 μV | 76 | 62 | 12.3 | 10.0 |

Please note that the SNRs and ENOBs are proportional to the ADC dynamic range for a given bandwidth, so the SNRs in decibels for 5 V and 1 V ADC full scales differ by $20 \cdot \log_{10}(5/1) \cong 14$ dB and ENOBs by $\log_2(5/1) \cong 2.3$ bits.

The numbers in Table 1 should be considered as "asymptotic ones" and most likely we will never see anything like this in real datasheets. However, at least theoretically, if the noise is defined by the terminating resistor, one could get even better numbers. For example, we could cool the 50 Ω resistor or replace it by an active circuit that provides the correct voltage to current ratio defining the impedance at the end of the coaxial cable, but produces less noise than a physical resistor.

More realistic numbers are listed in Table 2, which were calculated in a similar way as Table 1, but assuming that all the noise related to the ADC operation is 10 times higher, so it comes from $50\ \Omega \cdot 10^{\,2} = 5$ kΩ resistor, which, as shown later, better reflects the noise performance of real ADCs. Please note that these numbers are meant only to give some rough estimates. As the noise voltages in Tables 1 and 2 are different by factor 10, the corresponding SNRs differ by $20 \cdot \log_{10}(10) = 20$ dB and the ENOBs by $\log_2(10) \cong 3.3$ bits.

Please note the noise, SNRs and ENOBs that one sees in both tables for the largest system bandwidth of 1 GHz. This is why ADCs with 1 GHz analog bandwidth (and therefore oscilloscopes) are not likely to ever achieve true 16-bit resolution. Consequently, designers of beam instrumentation systems with GHz bandwidths may have hard time to get dynamic ranges close to 100 dB. At the time of writing this paper the noise performance of the best commercial ADCs with 1 GHz analog bandwidth is close to the numbers in Table 2 for 1 V dynamic range, with ENOBs in the order of 10 and SNRs about 60 dB.

## 3.4 Clock jitter

The time instants at which the ADC input signal is sampled are defined by the sampling clock. In the example shown in Fig. 10(a), the sampling instants are set by the rising edges of the clock at the level of 50 % of its amplitude. In a perfect case, the sampling instants are spaced by the same sampling period $T_s$. In reality, however, the sampling instants are not perfectly equidistant, but the sampling jitters around its nominal values, for example due to noise present in the clock signal. As illustrated in Fig. 10(b), during the time corresponding to the sampling jitter $\Delta t$ the input signal changes by $\Delta s$ and

$$\Delta s = \frac{ds}{dt} \Delta t \qquad (24)$$

Now let's consider a sinusoidal ADC input signal of the form

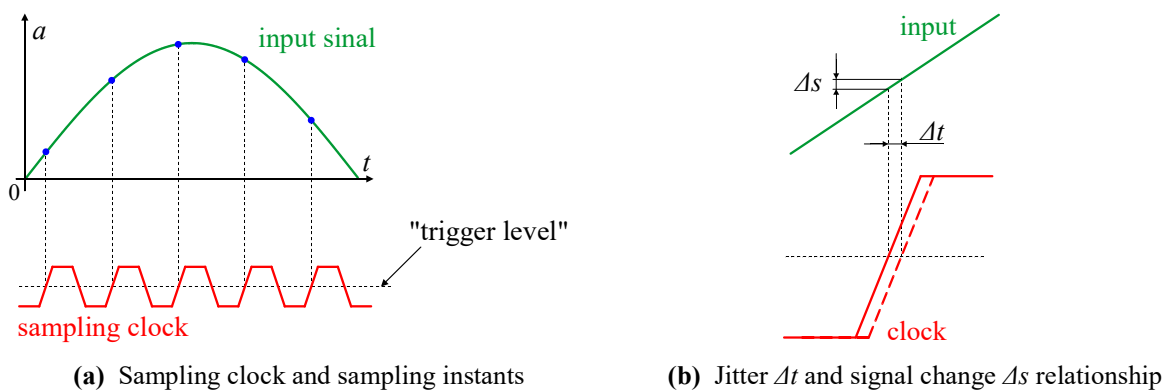

(a) Sampling clock and sampling instants  (b) Jitter $\Delta t$ and signal change $\Delta s$ relationship

**Fig. 10:** Illustration of the clock jitter concept. Figure (a) shows a clock signal defining sampling instants, while Fig. (b) explains the relationship between the sampling jitter $\Delta t$ and the corresponding signal change Δs.

$$s = A \sin(2\pi f t) \tag{25}$$

and we put a classic assumption that the worst-case signal amplitude change due to clock jitter $\Delta t$ should be smaller than 0.5 LSB. Taking into account (1) this requirement can be written as an inequality

$$\max\left(\frac{ds}{dt}\right)\Delta t < \frac{2A}{2^n} \tag{26}$$

where $n$ is the number of bits of the ADC. Please note that the maximal signal derivative is in fact the slew rate (20). Then the condition seen in many books for the allowable clock jitter $\Delta t$ that does not yet ruin the ADC performance is

$$\Delta t < \frac{1}{2\pi f 2^n} \tag{27}$$

Table 3 lists the clock jitter limits for typical $n$ values and three frequencies differing by three orders of magnitude.

At the time of writing this paper the best dedicated chips producing ADC clocks can achieve jitters in the order of 50 fs. To give some feeling of this number let's imagine that the ADC clock signal is transmitted from a source with no jitter over a typical 1 m coaxial cable. As the cable would have delay of some 5 ns, then 50 fs corresponds to 10 μm of the cable length. The 50 fs jitter would be caused by the cable if its length changes randomly by 10 μm.

It can be seen that in Table 3 there are clock jitters smaller than 50 fs. This means that with the current standard clock technology such numbers would be difficult to achieve and the ADC performance for such cases would be limited by the achievable clock jitter. It is interesting to see that such problems are already present for 1 MHz sampling with a 24-bit resolution and 1 GHz sampling with a 12-bit resolution.

**Table 3:** Maximal allowed clock jitters corresponding to noise smaller than 0.5 LSB for different combinations of the signal frequency and ADC resolution.

| Signal frequency | 8 bits | 12 bits | 16 bits | 24 bits |
|---|---|---|---|---|
| 1 kHz | 620 ns | 39 ns | 2.4 ns | 9.5 ps |
| 1 MHz | 620 ps | 39 ps | 2.4 ps | 9.5 fs |
| 1 GHz | 620 fs | 39 fs | 2.4 fs | 9.5 as |

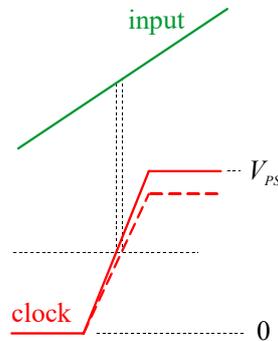

**Fig. 11:** Explanation how the quality of the supply voltage ($V_{PS}$) powering clock circuitry can influence the active slope of the sampling clock and in consequence the clock jitter.

Fortunately, there are ADC architectures, like delta-sigma (ΔΣ), for which the jitter requirements are relaxed with respect to (27), due to the fact that in such ADCs the actual sampling process is much faster than the ADC data rate and a special data treatment is involved. The related theory is nevertheless quite complex and there is no simple formula to calculate the required jitter, as it depends on many details of the actual internal ADC signal processing.

Generally, digital clock signals for high-performance ADCs should be treated as sensitive analog signals and thus they should not come directly from complex digital chips, like FPGAs. The most important reason is explained in Fig. 11. Imagine that the clock in the example comes from CMOS logic, where the digital states are essentially ground (0) and the power supply ($V_{PS}$, for example 3.3 V). For very fast clocks the "active clock slope" is an important fraction of the clock period, so a considerable part of the noise from the power supply node, defining the high logic state, is converted into clock jitter. Of course, the faster the slope, the smaller the "conversion factor" from the voltage noise to the clock jitter.

In real fast clock circuitry one uses differential signals to minimise the presented effect, which is likely to appear on the ADC differential clock input as a common mode signal. Also, adequate fast logic families can be used with "stabilised" logic states, which are therefore less dependent on the noise present on the power supply voltage. The output levels of modern FPGAs can also be configured to form differential outputs with amplitudes according to one of the logic standards with a reduced dependence on the power supply quality.

On the electronic market there are many specialised integrated circuits for producing ADC clocks of very high quality. Most often they contain very low jitter clock generators that can be synchronised to external sources, which have much relaxed requirements for the clock jitter. Typically, one such a chip is used to clean up ADC clocks coming from an FPGA.

## 4 Signal sampling

In the sampling process the continuous ADC input signal gets converted into a sequence of numbers, with all the signal between the "sampling points" being completely lost. It is remarkable that under quite reasonable conditions one can fully reconstruct the continuous input signal from its samples. In the simplest way the conditions can be abbreviated to a simple theorem, called the sampling theorem, which can be expressed as

$$\text{sampling frequency} > 2\, f_{max} \tag{28}$$

where $f_{max}$ is the highest frequency represented in the sampled signal.

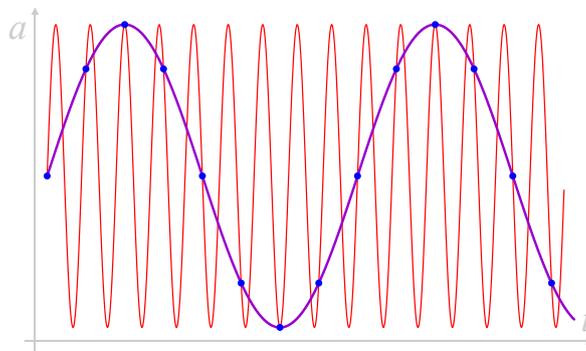

**Fig. 12:** An example of signal sampling with aliasing, when the sampling theorem (28) is not satisfied and the signal reconstructed from the samples (violet) is different from its sampled original (red).

An illustration of the sampling theorem is shown in Fig. 12, where condition (28) is not satisfied. What happens in the example is called aliasing. In such cases the signal reconstructed from the samples is different from the analog original. A typical method to guarantee that no aliasing occurs is to place a low-pass filter before the ADC with its cut-off below half of the sampling frequency. In such cases the filter is called an anti-aliasing filter. Please note that the filter should assure that no spectral content can pass at half of the sampling frequency, so some room for developing the filter attenuation is needed. For certain ADC architectures, like delta-sigma, where the actual sampling takes place at a much higher frequency than the output data rate, the requirements for the frequency characteristic of the anti-aliasing filters are much relaxed, making the filters easy and cheap to implement.

For some applications the aliasing is used explicitly for shifting the frequency content of a signal. In such cases the ADC is used in a similar way as a frequency mixer, typically to lower the frequency content of a signal. Please note that this is exactly the case in the "aliasing example" of Fig. 12. Some ADCs are dedicated for frequency conversion applications and in such cases you can often see in their datasheet that the allowable highest input frequency is much higher than the fastest sampling rate. An example is LTC2204 with the maximal input frequency of 700 MHz and the sampling rate of 40 MS/s.

For the sake of simplicity in this paper we assume that the sampling process is infinitely fast, while for real ADCs this is of course not possible. This has some consequences, however, most often of a higher order. Another simplification assumed in the paper is that the output data appears on the ADC digital output immediately, neglecting the internal processing time. Depending on the ADC architecture, this delay can be quite important and most often is expressed in units of the ADC sampling periods. For some ADCs this delay can be even a few tens of sampling periods. Such delays introduced by the ADC may be very important if the data is used in a feedback system. If the ADC delay is significant with respect to the feedback speed, then the delay may affect the stability of the feedback.

## 4.1 Digitising sinusoidal signals

In many beam instrumentation systems the information about beam parameters is carried by sinusoidal signals. The signals may have this form because of the nature of the beam sensor, for example an RF cavity of a beam position monitor, or because the signals underwent narrowband filtering using band-pass filters, while originally they were, for instance, very short pulses.

A typical way of digitising a sinusoidal signals is called IQ demodulation. In this technique the sampling frequency is exactly four times the frequency of the sampled sinusoid, as shown in the example in Fig. 13 for two cases with different phase relationship between the sampling and the input signal. The ADC data is grouped in two data streams, each formed by taking every second sample. The streams are called traditionally $I$ (from "in phase") and $Q$ (from "quadrature"). The digitised signal can then be represented by a vector with amplitude $a$ and phase $\varphi$ and both can be calculated from the $I$ and $Q$ data as

$$a = \sqrt{I^2 + Q^2} \tag{29a}$$

$$\varphi = \arctan\frac{Q}{I} \tag{29b}$$

where the phase is measured with respect to the sampling clock phase. Taking this into account the phase relationship between more signals can be evaluated.

The $I$ and $Q$ parts are in fact sampled in the "boundary case" of the sampling theorem (28) with exactly two samples per period. This results in constant $I$ and $Q$ values when the amplitude and phase of the input signal stay unchanged. Consequently, the $I$ and $Q$ parts can be low-pass filtered in the digital domain, potentially allowing very precise measurements at the expense of reduced measurement bandwidth.

A typical challenge of IQ demodulation in beam instrumentation systems is to assure a good synchronism between the digitised signal and the ADC sampling clock, especially if phase measurements are involved. Often the sampling clock is derived from the accelerator RF system and is harmonically related to the beam revolution frequency $f_{rev}$. During beam acceleration (or deceleration in some machines) $f_{rev}$ varies and the ADC clock should follow it. For some accelerators the $f_{rev}$ changes so much that the ratio between its maximal and minimal value can be quite large. For example, in CERN Antiproton Decelerator (AD) this ratio is about 10 and in such a case producing an ADC clock following the $f_{rev}$ changes is quite a challenge.

To explain why one has synchronisation difficulties when the ADC sampling frequency changes, let's consider a clock signal with frequency $f_{clk}$, which is delivered to an ADC from the RF system as a sinusoidal signal through a cable with delay $\tau_{clk}$

$$s_{clk} = \sin(2\pi f_{clk}(t + \tau_{clk})) \qquad (30)$$

The ADC digitises our beam instrumentation sinusoidal signal and from the *I* and *Q* data we calculate its phase with respect to the sampling clock. During beam acceleration the machine revolution frequency changes and so does the sampling clock frequency. For a clock frequency change $\Delta f_{clk}$ clock signal $s_{clk}$ (30) becomes

$$s_{clk}' = \sin(2\pi (f_{clk} + \Delta f_{clk})(t + \tau_d)) \qquad (31)$$

Thus, even if our beam instrumentation signal has exactly the same phase with respect to the circulating beam and the RF system, the phase of the ADC clock changes and the phase difference between $s_{clk}$ and $s_{clk}'$ calculated for $t = 0$ is

$$\Delta\varphi_{clk} = 2\pi(f_{clk} + \Delta_{clk})\tau_d - 2\pi f_{clk}\tau_d = 2\pi \Delta f_{clk}\tau_d \qquad (32)$$

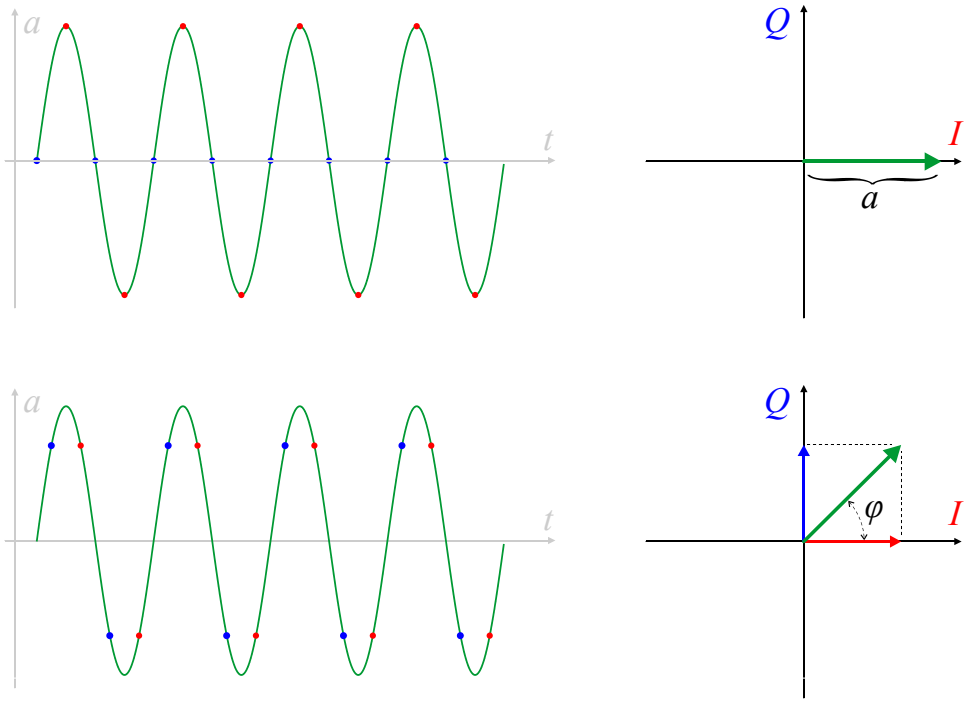

**Fig. 13:** Illustrations of IQ demodulation for zero (top) and 45° (bottom) phase relationship between the sampled signal and the sampling clock.

When the sampling frequency changes, the delay in the clock distribution network gets converted into a change in the sampling phase. Fortunately, in many beam instrumentation systems one measures only amplitudes and then the effect can often be neglected when using IQ demodulation.

### 4.2 Digitising pulse signals

As described before, digitising sinusoidal signals is very inexpensive in terms of the required sampling rate, since only four samples per signal period are required, which is just twice the theoretical limit (28). Unfortunately, producing sinewaves often involves narrow-band filtering, which decreases time resolution of the measurements. This is why in many beam instrumentation systems the ADCs digitise pulse signals. Often each signal pulse corresponds to one beam bunch and the beam instrumentation system is required to provide measurements for each bunch separately.

In beam instrumentation systems with pulse signals one has two basic options for the ADC sampling clock:
- a "synchronous clock" that is synchronised to the circulating beam;
- an "asynchronous clock" with a constant frequency, unrelated to the circulating beam.

The option of a synchronous ADC clock for pulsed signals is very similar to the clock described earlier for IQ demodulation. The synchronism between the ADC input signal and the ADC clock helps in systems for which the measured quantity is derived from a signal that changes from one machine turn to another. A classical example is a tune measurement system, where the machine tune is derived from turn-by-turn changes of the beam position.

Synchronous ADC clocks are beneficial for a good representation of the digitised signals with a relatively small number of samples. In the extreme case, when one is interested only in the amplitude of the pulse, the amplitude can be measured with just one sample if the sampling phase is sufficiently well adjusted, as shown in Fig. 14(a). However, if the sampling phase changes, for example due to the change of the revolution frequency combined with the delay in the clock distribution network (32), the measured amplitude will change as illustrated in Fig. 14(b). A way of avoiding such issues is to fit a model signal shape to the limited number of samples, allowing a better maximum evaluation.

The largest advantage of using an asynchronous ADC clock is its simplicity. Contrary to the synchronous clock, it can be generated locally close to the ADC board or even on the board itself, so there is no need for a distribution system. With large machines a dedicated distribution network for beam synchronous timing can be quite complex and expensive. Therefore, if a system can operate with locally generated clocks, this is a preferable approach, especially for large, distributed beam instrumentation systems, like beam position monitors.

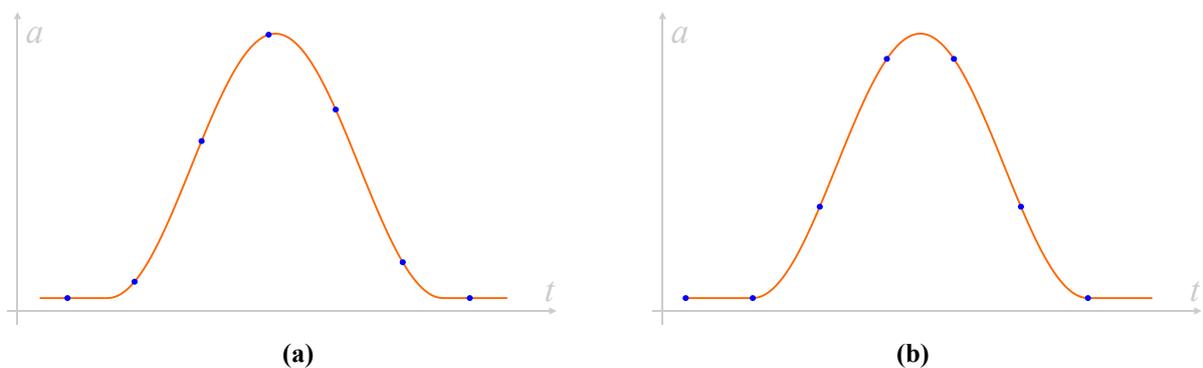

**Fig. 14:** Example of pulse signal sampling for two phase relationships between the signal and the sampling clock.

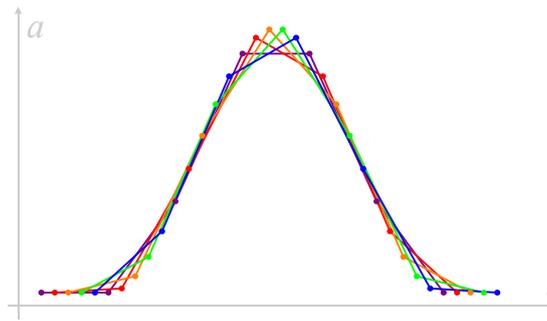

**Fig. 15:** Asynchronous sampling results in random sampling phase with respect to the circulating beam, so reconstructing the pulse from the samples is not straightforward with a limited number of samples.

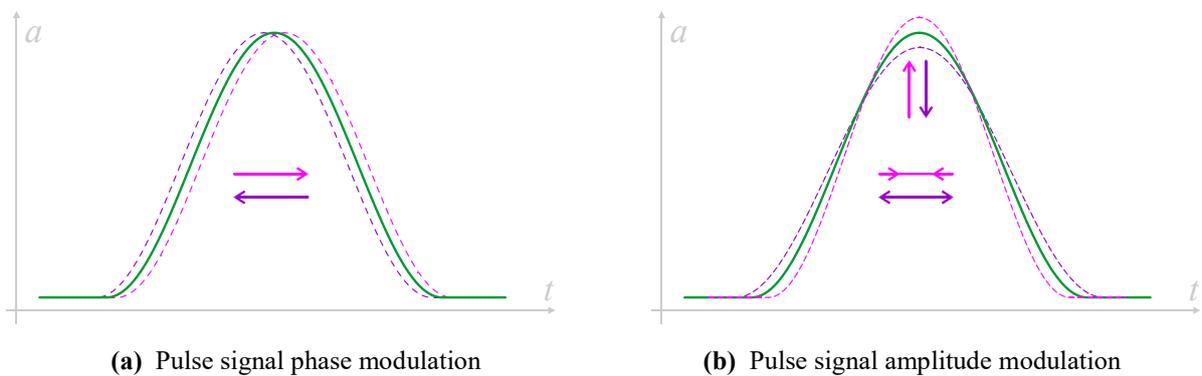

**(a)** Pulse signal phase modulation    **(b)** Pulse signal amplitude modulation

**Fig. 16:** Examples of effects caused by beam synchrotron motion. (a) Phase modulation of the signal pulse. (b) Amplitude modulation of the beam pulse, coupled to the corresponding bunch length changes, as the bunch charge (integral) is conserved.

Asynchronous ADC clocks with a constant frequency can be generated with relatively simple circuitry and allow achieving the best clock quality with the smallest jitter. Furthermore, modern very fast ADCs have quite complex internal processing and data buses with error correction, which often just do not accept clock frequency changes. This is why most often very fast ADCs work with sampling clock asynchronous to the circulating beam.

Asynchronous clocks result in a random phase relationship between the sampling instants and the beam signal, as illustrated in Fig. 15. The effect can be minimised by increasing the sampling rate or by stretching the beam pulses by using adequate low-pass filtering. Please note that the second option can be very efficient and inexpensive, especially if the distance between the adjacent pulses is much larger than the length of the pulses. Making pulses longer lowers their amplitude, but it is often much more efficient to amplify the filtered signal than to increase the sampling rate. In some systems the beam signals are anyway too large to be processed and digitised directly so, especially in such cases, it is worth considering low-pass filtering, instead of just attenuation. Then more samples per signal pulse can come almost for free, without increasing the ADC sampling rate, just at the expense of the low-pass filters.

Digitising pulse signals is in general more difficult than digitising sinusoidal signals. A sinusoid is a well-defined shape, which unfortunately is not the case for pulse signals, as their shapes can vary a lot. One should not forget that signals induced by circulating beam can change quite a bit from one turn to another, for example due to beam synchrotron motion. Synchrotron motion can make the beam pulses phase-modulated, as illustrated in Fig. 16(a), and amplitude-modulated, as shown in Fig. 16(b). Such effects should be taken into account when choosing the ADC sampling rate. If needed, the modulations can be at least partially supressed by an adequate low-pass filtering of the beam pulses.

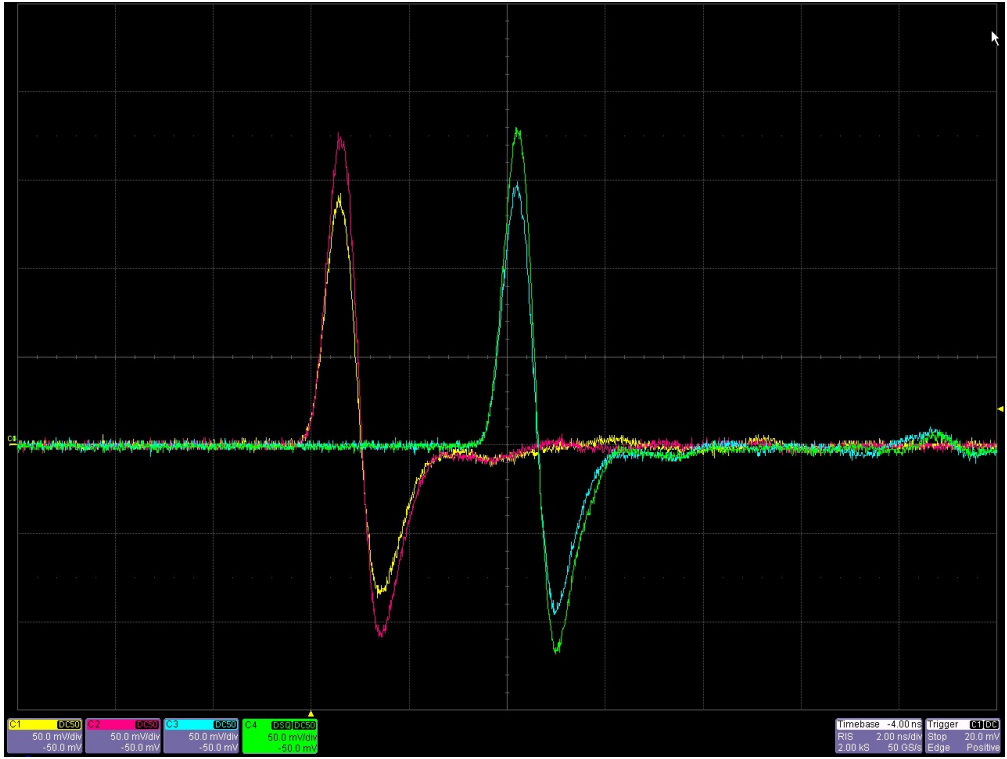

**Fig. 17:** Oscilloscope measurements of the signals from two pairs of 10 mm button electrodes, CERN SPS beam. Amplitude axis: 50 mV/div, time axis: 2 ns/div, equivalent 50 GS/s in RIS mode with 2.5 GS/s sampling, 1 GHz analog bandwidth.

When designing a beam instrumentation system and its acquisition it is always good practice to make some measurements of the raw signals from your beam sensors, to be sure what they look like. Often simple measurements can be more useful than even the most sophisticated simulations, which may not include all details and imperfections that are important in the real system. An example of beam signals from two pairs of button electrodes induced by one bunch circulating in the CERN Super Proton Synchrotron (SPS) is shown in Fig. 17. The signals were measured with an oscilloscope sampling at 2.5 GS/s but run in the random interleaving sampling (RIS) mode, resulting in the equivalent sampling rate of 50 GS/s and giving very fine time resolution of the measurement. The analog bandwidth of the oscilloscope was 1 GHz, with the corresponding rise time of about 0.3 ns. We can see that the measured signals have rise times in the same ballpark, so the speed of the oscilloscope may actually have limited the speed of the observed signals. The signals from each electrode pair are separated by some 3.5 ns, corresponding to the beam travel time between the electrode pairs.

What is important for the matters discussed in this paper is that even if the circulating bunch has the time profile similar to a Gaussian shape, the corresponding electrode signals are heavily differentiated. One can also see that after the pulses the signals do not go to zero, but there are some undulations lasting much longer than the bunch length. Most likely they are related to some signal reflections between the cables and the electrodes. Observing details of beam signals may help in choosing an adequate analog signal processing and good digitisation parameters for the designed beam instrumentation system.

# 5     What sampling rate is needed?

Choosing a good sampling rate for ADCs in a beam instrumentation system is probably the most important and difficult decision one has to make when starting a new project, especially if high frequency signals are involved, which potentially require very fast sampling rates. In such difficult cases selection of the sampling rate is often an iterative process, in which one evaluates the options and trade-offs, for example the trade-off between sampling speed and resolution, sampling speed and cost, or sampling speed and the complexity of the analog signal processing prior to the ADC. However, the most important input for the sampling speed selection is the time resolution of the measured quantity stated in the specification of the beam instrumentation system to be developed. To illustrate the importance of the required measurement time resolution let's look at three examples of real and operational systems measuring beam positions in the LHC. Despite the fact that all three systems are based on very similar beam signals, the ADCs used have sampling rates differing by almost six orders of magnitude for the most extreme cases. The systems are built this way because they have completely different requirements for the measurement time resolution. The following examples are focussed on the beam signal sampling and other aspects of the systems are here much simplified, including "idealisation" of the depicted signal shapes.

The first system is the so-called "head-tail monitor", which is mostly used to study very fast beam instabilities happening inside the circulating bunches. The system is quite simple in its principle being based on almost direct acquisition of the beam signals coming from large-bandwidth beam position monitors by using fast oscilloscopes. A sketch of the time structure of the digitised signals is shown in Fig. 18(a). Each signal consists of beam pulses some 2 ns long and spaced every 25 ns. The system provides beam position changes inside the short bunches, so the analog bandwidth of the oscilloscope is 4 GHz and the sampling is done at the rate of 10 GS/s. This results in some 20 samples per beam pulse, which are asynchronous to the beam. The sampling ticks are represented on the sketch by red vertical lines.

As a consequence of the very high sampling rate, the acquisition of the "head-tail monitor" is limited to 10-bit resolution, with only 7.2 ENOB, which was still one of the best options available on the market at the time of building the system. The oscilloscope that is used costed a fortune, but as the system consists of only one dual-plane BPM per LHC beam, there are two 4-channel oscilloscopes in the whole system. The system provides unique measurements to LHC operators, so the invested money is justified. However, such expensive oscilloscopes are not an option for acquiring signals from the over one thousand LHC BPMs. Furthermore, the LHC head-tail monitor produces some 890 000 samples per LHC turn, resulting in the data rate of 12.5 GB/s per oscilloscope channel. The whole system with eight channels produces then data at a rate of 100 GB/s and only a small portion of the data can be stored and analysed. This is another reason why such fast digitisation of BPM signals is only practical to use on two dedicated LHC BPMs.

The LHC head-tail monitor is a good example of a "brute force" fast digitisation of beam instrumentation signals, with all of its typical consequences: limited resolution, very high data rates and a large cost. In this case the cost is kept reasonable by limiting the size of the system and other drawbacks are justified by the required time resolution of the system.

The subject of the second example is the standard electronics used to process the signals from the over one thousand LHC BPMs that were already mentioned. The signals from the BPMs are very similar to those used by the head-tail monitor and this system is required to provide a beam position for each circulating bunch, so one measurement every 25 ns. The system is very large, so the cost per BPM was a major optimisation factor. This is why the sampling rate was chosen to be as slow as possible, that is every 25 ns, corresponding to a 40 MHz measurement rate. A very simplified sketch of the signal processing is shown in Fig. 18(b). The first important operation is passing the 2 ns beam pulses through 70 MHz low-pass filters in order to stretch the pulses to some 10 ns long and make then much easier to deal with. After quite complex analog processing at the end the signals are integrated and sampled once the integral reaches its steady state during the "silence gap" between two adjacent pulses. This way the amplitude of the initial 2 ns signal can be measured with just one ADC sample. The LHC system was developed and built some years ago, so the ADCs used only have a 10-bit resolution with some 9.6 ENOB, which was a very good performance at that time. The LHC BPM electronics measures the beam position for each bunch separately, so it must tag the bunches. For this the BPM system uses a 40 MHz signal synchronous to the beam that is distributed optically all over the LHC. The BPM system provides some 3600 samples per LHC turn, resulting in the data rate of some 100 MB/s per dual-plane BPM. As there are over one thousand BPMs in the whole system, this still results in a lot of data.

The LHC BPM system is a good example of a large, universal beam instrumentation system, in which clever analog processing of the beam signals allowed reducing the required sampling frequency to the minimum defined by the specified measurement rate of the system. This approach allowed optimisation of the system cost and amount of generated data, which has to be post-processed and often continuously logged.

In the last example the used analog processing allows the sampling rate to be still further lowered, because the system is required only to provide measurements once per beam revolution, as shown in the signal sketch in Fig. 18(c). The beam signals come from LHC BPMs and they are again very similar to the signals of the head-tail and LHC BPM systems. The analog processing starts with low-pass filtering

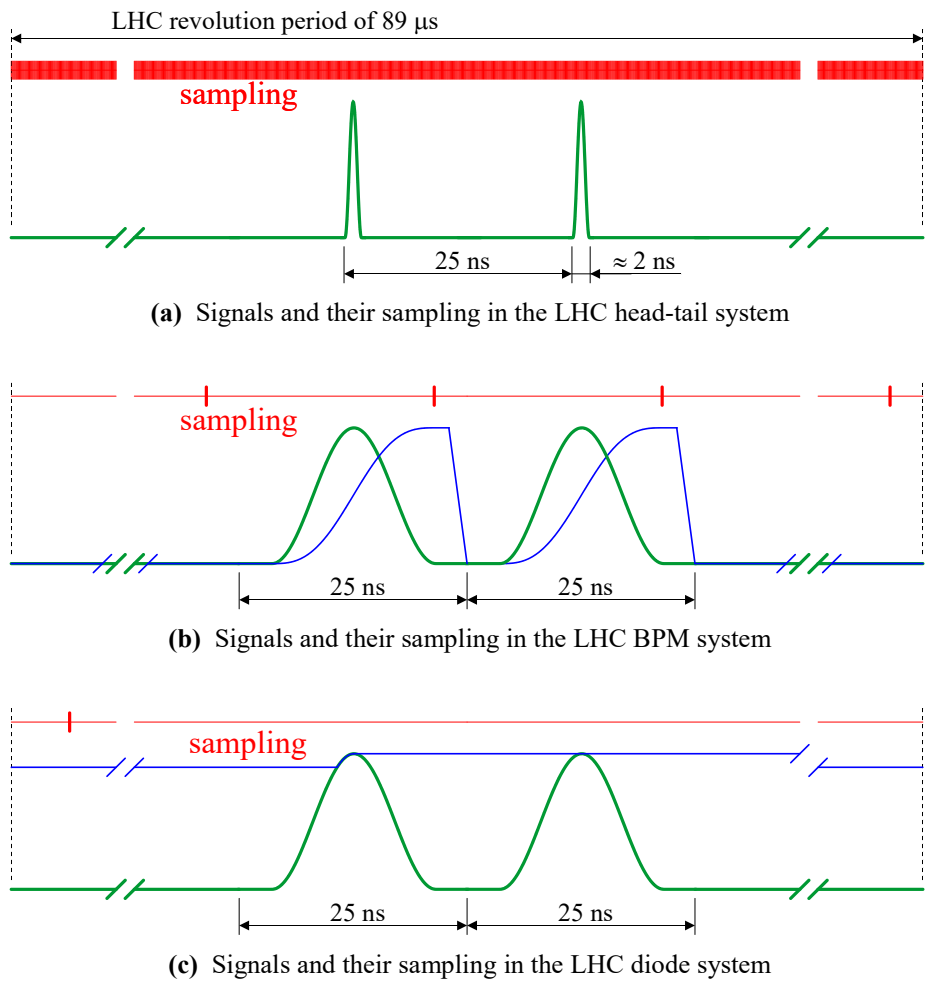

(a) Signals and their sampling in the LHC head-tail system

(b) Signals and their sampling in the LHC BPM system

(c) Signals and their sampling in the LHC diode system

**Fig. 18:** Examples of idealised beam signals, their simplified analog processing and sampling for three LHC systems based on very similar beam sensors. As required measurement time resolution is quite different for each system, they all have dissimilar analog signal processing, demanding completely distinct sampling rates.

to 80 MHz and finishes with diode peak detectors. The detectors "sample" the beam pulses close to their maxima and the corresponding voltages are "memorised" on parallel RC circuits, with a large time constant, much longer than one LHC revolution. This trick allows sampling just once per LHC revolution turn, at a 11.2 kHz rate, which makes it possible to use 24 bit ADCs with an ENOB of about 18. The diode BPM electronics provides output data at a rate of some 150 kB/s per dual-plane BPM.

The presented diode system illustrates the power of smart analog signal processing, allowing sampling 2 ns beam pulses only once per 89 µs, giving the "beam pulse stretching ratio" of some 45 000. All three examples are meant to support the conclusion that the goal sampling rate in a beam instrumentation system primarily depends on the specification of the system. In general, the sampling frequency can be as low as the measurement rate that the system is required to provide. Such "ultimately low sampling" may however require complex analog processing, which often can be simplified if one accepts faster sampling. Such discussion on the sampling rate could be continued here for quite a few more pages, but at the end it is very much a compromise between many factors, like resolution, cost, complexity of the analog processing, amount of produced data, available know-how, manpower and often the experience of people involved. For sure the specification of the system can be fulfilled in many ways. Nevertheless, in most cases it is worth starting from simpler ideas and go to more complex ones only if necessary.

The simplest approach, which is probably the dream of all young beam instrumentalists, is connecting the ADC almost directly to the beam sensor. If your signals are not too difficult, that is they are not too small and, above all, not too fast for the current state of the ADC technology, this approach may actually work, especially if you use adequate filters, amplifiers and attenuators to adapt the signals to your ADC. If you can find and afford an ADC, which allows such a simple approach while fulfilling all the system requirements, then this is probably the first option to think of and evaluate. With continuous improvements in the ADC technology such an approach should be possible for faster and faster signals.

If the simplest approach does not satisfy the system specification, you should then consider some analog signal processing, which helps to make your signals easier to handle. In most cases this just means converting fast signals into something slower, which can be handled by ADCs you can use. Again, it is probably best to start from simpler options and once you have a clear idea of what can fulfil the system requirements, build a prototype of this analog processing and test it together with the selected ADC.

## 6   How many bits are needed?

The question of how many bits are required for a beam instrumentation system is by far easier to answer than the previous one on the required sampling rate. How this can be done is illustrated with the following example.

Let's imagine that we are about to build a system measuring the number of charges circulating in one of the LHC beams. This can be done by measuring the beam current and calculating its one-turn integral. Our beam signal comes from a DC beam current transformer, but as far as the required number of bits is concerned, we even do not need to know exactly what the signal looks like.

The system should measure the smallest beam intensity of $Q_{min} = 5 \times 10^9$ elementary charges (a so-called pilot beam) with 1 % resolution. This results in $Q_{LSB} = 0.01\ Q_{min} = 5 \times 10^7$ charges, which defines the charge corresponding to the ADC least significant bit above the noise. In general, the ADC resolution in a good beam instrumentation system should be much smaller than the noise coming from the beam sensor and its electronics. This assumption assures that the acquisition part of the beam instrumentation system does not limit the system performance.

The system should measure nominal bunches up to an intensity of $Q_n = 3\times10^{11}$ charges and up to maximum of $n_{max} = 3000$ bunches, which results in the maximal measured charge $Q_{max} = n_{max} Q_n = 9\times10^{14}$. The dynamic range of the acquisition system should be then $D = Q_{max}/Q_{LSB} = 1.8 \times 10^7$ and the required number of bits is therefore $R = \log_2(D) = 24.1$.

As shown, calculating the required resolution is often quite easy. What is more difficult though is to build a system with the calculated resolution. In the case of our example the required 24.1 bits suggests for us to use a 24-bit ADC sampling once per LHC revolution turn, that is at $f_s = 11.2$ kHz. Indeed, today one can find excellent 24-bit ADCs sampling at this rate, however, they have more like 18 effective number of bits, leaving us with 6 missing bits.

The simplest way of "gaining" these 6 bits is to average the signal by factor $g = (2^6)^2 = 4096$, assuming that the ADC resolution is limited by its white noise. This would reduce the measurement rate to $f_m = f_s / g = 2.7$ Hz. If such a measurement rate is acceptable, then this is the easiest way to achieve the required resolution. If needed, the measurement rate can be increased by using a moving average or another digital filter, which would have a similar bandwidth, but would allow a higher measurement rate.

The previous illustration with ADC sample averaging is a classic example how one can increase the system resolution by limiting the system bandwidth. In the example the bandwidth was reduced in the digital domain by data filtering, but the system bandwidth can be also limited before the ADC by an adequate processing of the analog signals. Some examples of such processing were given in the previous chapter on the required sampling rates.

Nevertheless, in many cases we cannot decrease the bandwidth any further, as it is already reduced to the limit defined by the system specification, and still there is no ADC on the market that can do the job or which we can afford. Two options to overcome such situations are illustrated in Fig. 19.

The first option is to divide the dynamic range into two ranges. In the previous example it would mean that if we have small signals from the sensor, then we need to amplify them to the level corresponding to high signals from the sensor without the gain. Also it is possible to imagine a complementary scenario, where the small sensor signals are digitised "directly", while large signals need to be attenuated to make them fit to the ADC dynamic range.

The second option is to have many gain (or attenuation) ranges and to select the optimal one according to the actual level of the signal from the sensor, giving a "good level" on the ADC input. In the extreme case the system gain (or attenuation) can be changed continuously to always have the same level on the ADC input. If the system is built this way, the signal controlling the gain can be actually used as the measure of the sensor signal amplitude.

The options for dividing the system total dynamic range into two or more sub-ranges are conceptually simple, but in practice their realisation can be quite complex. It is especially the case if the particular gains have to be taken into account in the final measurement result. Another issue is what happens during the gain switching. Often during this process the system data is of reduced quality and cannot be used, resulting in gaps in the measurement, which may be very problematic, especially in protection and safety systems. Therefore, in many such systems gain switching is not an option.

One way to avoid gain switching is to have two or more "sub-systems" that run in parallel, with each one having different dynamic ranges, its own ADCs and data processing. Then one can select the data stream from the sub-system with the most optimal signal level. In many cases switching between the data streams can be made so fast that it is transparent for the users.

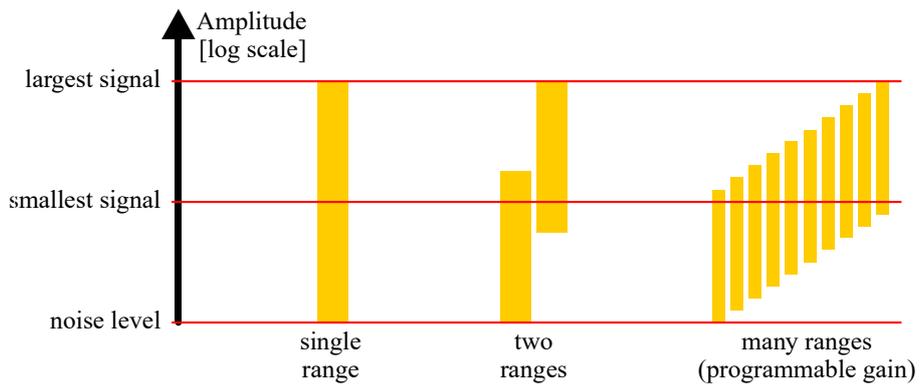

**Fig. 19:** Options for dynamic ranges: single range, two ranges and many ranges.

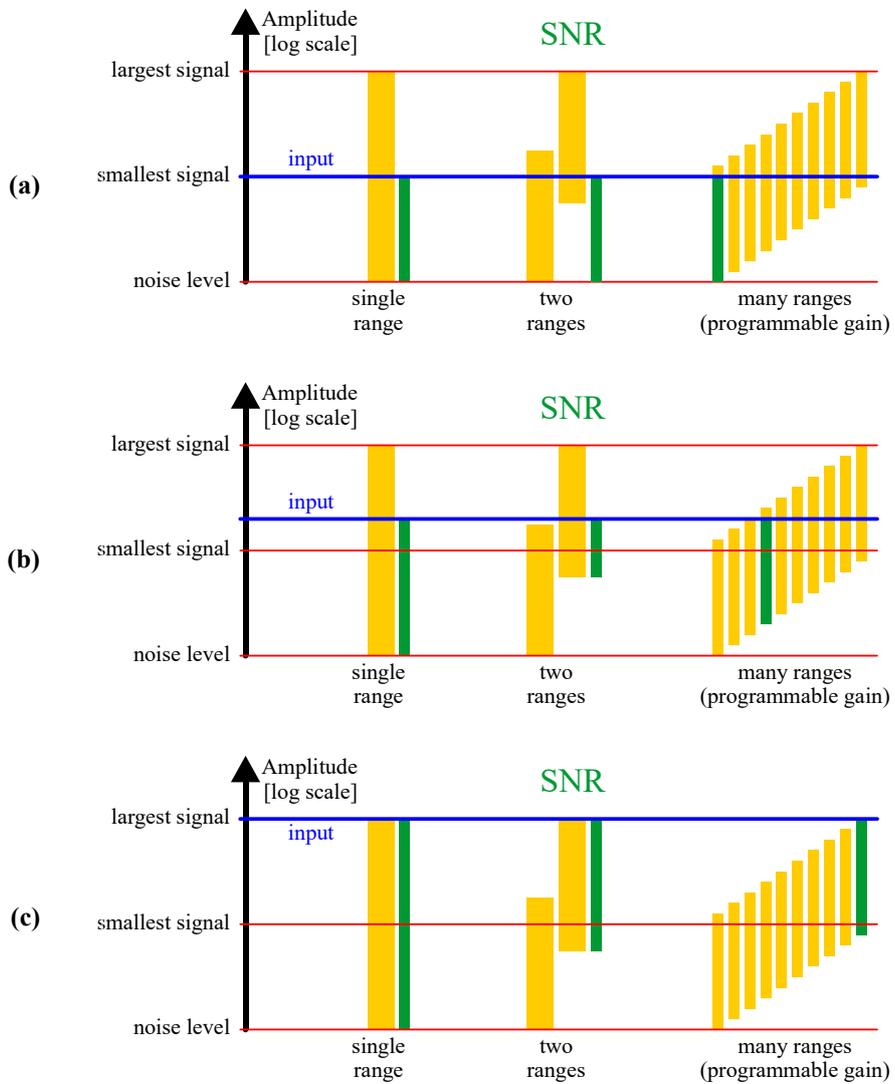

**Fig. 20:** Signal-to-noise ratios in systems with one, two and many dynamic ranges, with the input signal amplitude close to its minimum (a), above the threshold for the second range in the two-range system (b) and close to the signal maximum (c).

Another option is to make available to the users all of the data streams and leave it up to them to select the most convenient one. However, such solutions with parallel processing of beam signals in two or more dynamic ranges are complex and expensive and therefore they are typically reserved for very special applications, like safety systems.

Another consequence of using two or more dynamic ranges is the way how the system signal-to-noise ratio changes. The first example shown in Fig. 20(a) illustrates the case when the instrumentation signal is close to its minimum. In such conditions all three approaches, with single, two or many ranges, would give similar signal-to-noise ratios. If the signal increases, as illustrated in Fig. 20(b), then the signal-to-noise ratio in the single-range system increases. In the two-range system the signal-to-noise ratio may actually decrease if the signal level is such that it crosses into the second range. The signal-to-noise ratio on the many-range system stays more or less constant, as the system gain is regulated to maintain a good signal level at the ADC input. This is also the case, when the signal reaches the maximum of the dynamic range, as shown in Fig. 20(c). In such conditions the single-range system achieves its highest signal-to-noise ratio, while the two-range system has a similar signal-to-noise ratio as with the small signal in the previous range.

## 7   What is a nice ADC?

When selecting an ADC for a beam instrumentation system it is rare that only a single chip on the market is suitable for the job. If this is the case, then it is probably not a good sign, as it is likely that you are asking for very high-performance. In the majority of cases many chips can be considered, and then the question is how does one chose the most optimal one. This chapter contains a few hints, which may help in this process. The time that needs to be invested in choosing an ADC and making it work in a beam instrumentation system is often quite large, so selecting a "nice ADC", which has some chance to be used in other projects, is always a good idea. The suggestions included in this chapter mostly concern cases where the ADC is to be integrated into a dedicated board designed for a beam instrumentation system. However, even if one decides to buy a commercial module, the hints may help in choosing a module with a "nice ADC".

Let's start from a checklist of features that a nice ADC must have in order to be worth investing time and money:

- a good datasheet, with all important ADC parameters stated, together with a description of how they were measured; surprisingly, not all datasheets satisfy these elementary requirements;
- a good datasheet performance, which at least on paper fulfils the requirements of the beam instrumentation system, in preference with some margin;
- available for purchasing in the required quantities and with acceptable delays; it is a good idea to check the market before we select something that cannot be bought in the quantities we need;
- a differential analog input;
- the possibility to connect an external voltage reference; it is beneficial if the reference input is differential as well.

Once we have a short list of ADCs with the above features, we can check which ones have development kits and whether the kits can be used in the first prototype of our system. As explained later, a good development kit can simplify and speed-up the system prototyping. It is also a good idea to compare how one connects the ADC output bus to the external world. For very fast ADCs some interfaces can be quite complex and may complicate the reception of the ADC data.

Some ADCs belong to larger pin-compatible families with different sampling and resolution combinations. If we see that such an ADC can be used in our system, this may be an interesting option as without changing the board layout we could have versions suitable for other applications. Also, with a bit of luck, new members of the family could show up in the future and we could potentially upgrade our beam instrumentation system without redesigning the board. A good example of a pin-compatible family with "nice ADCs" is the LTC2204/5/6/7, with seven members having 16- or 14-bit resolution and sampling rates from 40 to 105 MS/s.

It could be also beneficial to use an ADC that has versions with different numbers of channels. Then, depending the needs, we could redesign the ADC board for a version with the most optimal number of channels. Such a redesign generally takes far less work than developing an ADC board from scratch. A good example of such a 24-bit family is ADS1271/4/8, having three members with one, four and eight channels.

Another way of looking at the term "a nice ADC" is from a practical perspective: a nice ADC is one that you can obtain with minimal manpower and money. This is where you can benefit from what is already in use around you. Once you have some idea for the ADC you need for your system, maybe try to look around and check what your colleagues use. With a lot of luck, you may find an ADC board, which fits well to your project. With less luck, you may find something that uses an ADC that is adequate for your project but that would require some redesigning, for example some adaptation to another digital bus, different board form factor, more ADC channels or different power supplies. Again, redesigning an existing board can be much easier and faster than starting from scratch.

# 8   ADC boards

An ADC is just an integrated chip and to use it in your system it must be put on a printed circuit board (PCB) with all of the circuitry it needs to operate, like power supplies, clocking, input amplifiers, a reference and output bus drivers. The board must be connected to something, which takes the ADC data and often provides some control to setup the ADC parameters by configuring its internal registers. For very fast ADCs producing high throughput data streams you may need to do some processing already on the ADC board to reduce the data throughput in order to make the following transmission and digital processing reasonable. Because of all this, you need an ADC board and the following are the most common ways to proceed:

A. use an existing ADC board that is already in use in another project or group (option "existing board");
B. buy a commercial board ("commercial board");
C. buy an ADC development kit and, if necessary, modify it for your project (option "dev-kit");
D. redesign an existing ADC board that is already in use in another project or group (option "redesign existing board");
E. design, prototype and produce a general-purpose ADC board in collaboration with other projects ("universal design in collaboration");
F. design, prototype and produce a custom ADC board for your project ("custom design").

Which option is preferable for your beam instrumentation system depends on many factors, like:
- how many boards you need;
- what is the probability that in the future you will need more such boards;
- how much experience with ADCs and PCB design you have in your team;
- how much time, manpower and money you have;
- what is the architecture of your beam instrumentation system;
- any special requirements your board has.

It is impossible to discuss in detail all of the above options and factors, which all should be considered when deciding on how to obtain ADC boards. Moreover, if you ask many people with experience and knowledge in the ADC and beam instrumentation matters for advice, they would likely not recommend you one scenario, as the final choice depends on so many fuzzy factors and sometimes also on personal experience and opinions. Therefore, in this paper you will only find some hints, which may help in making choices. Nevertheless, only you and your team can put all aspects into the context of your project.

If the option of an "existing board" can be considered, this is probably the first choice to study. In this case you can measure the board in question and verify experimentally that it fulfils all your requirements. On the other hand, the board should not exceed your requirements too much, as in such a case it may be expensive or inconvenient to use. This is however likely to matter only if you need a significant number of boards.

The option of a "commercial module" may be considered if you do not have the know-how to build your own board, you need only a few modules, or you do not have time for a dedicated development. Also, to use a commercial module your system has to be based on some industrial standards. While convenient for smaller projects, a commercial module may be a trap if you heavily underestimate the number of the boards you need and the boards are expensive. Then one day you may be upset to learn that, in addition to the ten boards that you already bought for a few thousand per piece, you need another hundred and now there is no time to develop a custom board.

Commercial ADC boards tend to accommodate a lot of functionality in order to make them more universal and suit broader spectrum of users. A common case is that they have also some DACs, which for many beam instrumentation projects are completely superfluous. Also, the very high component density that is typical for commercial modules does not help for ADC performance. This is why custom boards with larger dimensions may potentially have better performance. In general, denser PCBs are more difficult to design and to maintain a very good signal quality.

The "dev-kit" option may be very interesting if you need to evaluate an ADC, you need a few ADC boards for an educational project or you plan to develop your own board with the same ADC. This is actually how the author starts more or less all his projects. In a few cases ADC development kits were modified and installed in prototypes tested with beam. This allowed, in parallel with the beam tests meant to evaluate the analog electronics, to develop a custom board with the same ADC that was fully optimised for the developed beam instrumentation system.

The alternative "redesign an existing board" could be worth considering if you find an ADC board, which is "close enough" to what you need and you have access to all the documentation of the project. If you are lucky, you may need to modify the board only for mechanical dimensions, the output bus standard or required power supplies. Starting from an already functional board is by far easier, faster and cheaper than starting your custom design from scratch.

The option of a "universal design in collaboration" could be interesting if, for example, you need boards with high-speed ADCs and the board design looks laborious and expensive. Then if you find more clients so that you can produce more boards, the design cost per board can be smaller. In such cases experts from the involved projects can also work together to speed-up the development and prototyping. On the other hand, a "collaboration board" sometimes may be more complicated than what you actually need, as the board should satisfy more requirements and some of them may be more challenging than yours. Also, if you need only 10 boards and the collaboration project 1000, then it is not guaranteed that by collaborating you would gain time.

If your project is large enough, you need at least a few tens of ADC boards and you have the know-how and time, then you may consider a custom board for your project. In this case the board would be exactly as you like, with exactly all the circuitry you need, an optimal number of channels, clocking, power supply, dimensions and LEDs of your favourite colour. If you have designed one ADC board and saw all of the good sides of this approach, then most likely you next project will also be with a dedicated, custom ADC board. The most difficult decision is probably then to design your first ADC board.

It may be more difficult to decide whether to go for a custom ADC board if you need only a few boards. If the boards are not very high speed and you have time for a custom design, maybe you could consider using a development kit as the starting point and develop your custom board anyway. If you do not need anything too different to what the development kit uses and you have at least one electronics engineer in your team, then going this way is probably not a bad idea. If your ADC does not sample at a very high speed and your schematic is correct, then it is not likely that your ADC board will not work at all. What may happen, however, is that the ADC will not have the full datasheet performance due to limitations of the PCB design. On the other hand, if your ADC samples close to 1 GHz or higher, then it may happen that you cannot receive any reasonable data from the ADC, because of inadequate routing of critical high-speed digital traces. This is why designing boards with very high speed ADCs is by far more difficult and why a board sampling at a few GHz is in general not a good choice for learning PCB design. A better approach is probably to start from an easier ADC board and develop the related know-how in your team, so that if you need a more challenging ADC next time, you could do the design and the board would work right away.

In some beam instrumentation systems the ADC requirements are so special that you may not have any other options but making a custom design. For example, if you need an external ADC clock that changes quickly over a large range, you are unlikely to find a suitable of-the-shelf ADC board. Often custom ADC boards can achieve better performance than their commercial equivalents, because they do not contain any superfluous circuitry and they can have smaller component density. The freedom of choosing larger PCB dimensions may help in achieving better performance with custom ADC boards than with commercial boards having most often small industry standard dimensions.

Designing a PCB for a high-performance ADCs is quite a challenge because of the fact that one has to deal with sensitive analog signals in presence of many digital circuits, most of which are switching synchronously to the ADC clock. In addition to the internal switching circuits of the ADC itself, often there are external digital circuits receiving and processing the ADC data. All these circuits cause large pulse currents on the power supply lines, which in properly designed PCB are minimised by using many local storage capacitors. However, some residual pulse currents in the power distribution are inevitable, resulting in interference signals affecting the circuit ground planes and then propagating all over the PCB. The art of a good ADC PCB design is to minimise the influence of such interferences by an adequate usage of differential lines, power supply filtering and optimal circuit distribution on the PCB.

## 9   Summary

This paper, corresponding to a one-hour CAS lecture, describes selected aspects of analog to digital conversion and is aimed at helping in choosing an adequate ADC for a beam instrumentation system. An important part of the paper is devoted to discuss compromises between the ADC sampling rate, its resolution and the complexity of the analog signal processing before the ADC. Most important fundamental limitations of ADCs are also described, with the focus on aspects likely to affect the compromises faced during the selection of the ADC speed and resolution.

An overview of the slew rate, noise and jitter limitations are summarised in Fig. 21, along with noise performance of the ADCs mentioned in this paper (red crosses) and a few examples of ADCs with outstanding performance in 2019 (purple crosses). The noise performance is quantified in effective number of bits (left vertical axis) and decibel signal-to-noise ratio (right vertical axis), with both quantities related according to (7). The left axis spans over 13 bits, corresponding to $20 \cdot \log_{10}(2^{13}) \cong 78$ dB of the right axis signal-to-noise ratio, that is almost four orders of magnitude.

The two horizontal axes of the plot have different scales, allowing combining on one plot the quantities related to the sampling rate (the lower horizontal axis) and the system bandwidth (the upper horizontal axis). This however requires an assumption on the scale relationship.

Thus, it is assumed that the axes are related by the boundary case of the sampling theorem (28) and the sampling rate spans over six decades from 10 kHz to 10 GHz while the corresponding system bandwidth boundaries are twice lower, namely 5 kHz and 5 GHz.

The assumption that the ADC sampling rate is twice its analog bandwidth is sometimes a severe simplification, especially in case of ADCs foreseen for frequency conversion applications, like already mentioned LTC2204 that has the maximal sampling rate of 40 MHz and the analog bandwidth of 700 MHz. This is why LTC2204 is exceptionally marked on the plot with two crosses, one corresponding to the sampling rate (1a) and the second one to its analog bandwidth (1b). To keep the plot complexity reasonable, all of the remaining ADCs are indicated according to their maximal sampling rate, while their noise performance is specified, as close as the datasheets permit, for the half of this rate.

Many ADCs have more than one operation mode with different trade-offs between the sampling speed, signal-to-noise ratio and power. An example of such an ADC is the already mentioned ADS1278, which has four operation modes marked on the plot, namely high-speed (HS, 2a), high-resolution (HR, 2b), low-power (LP, 2c) and low-speed (LS, 2d). Again, to keep the plot clear, the remaining ADCs with more than one operational mode are marked only once, with the performance corresponding to the fastest mode.

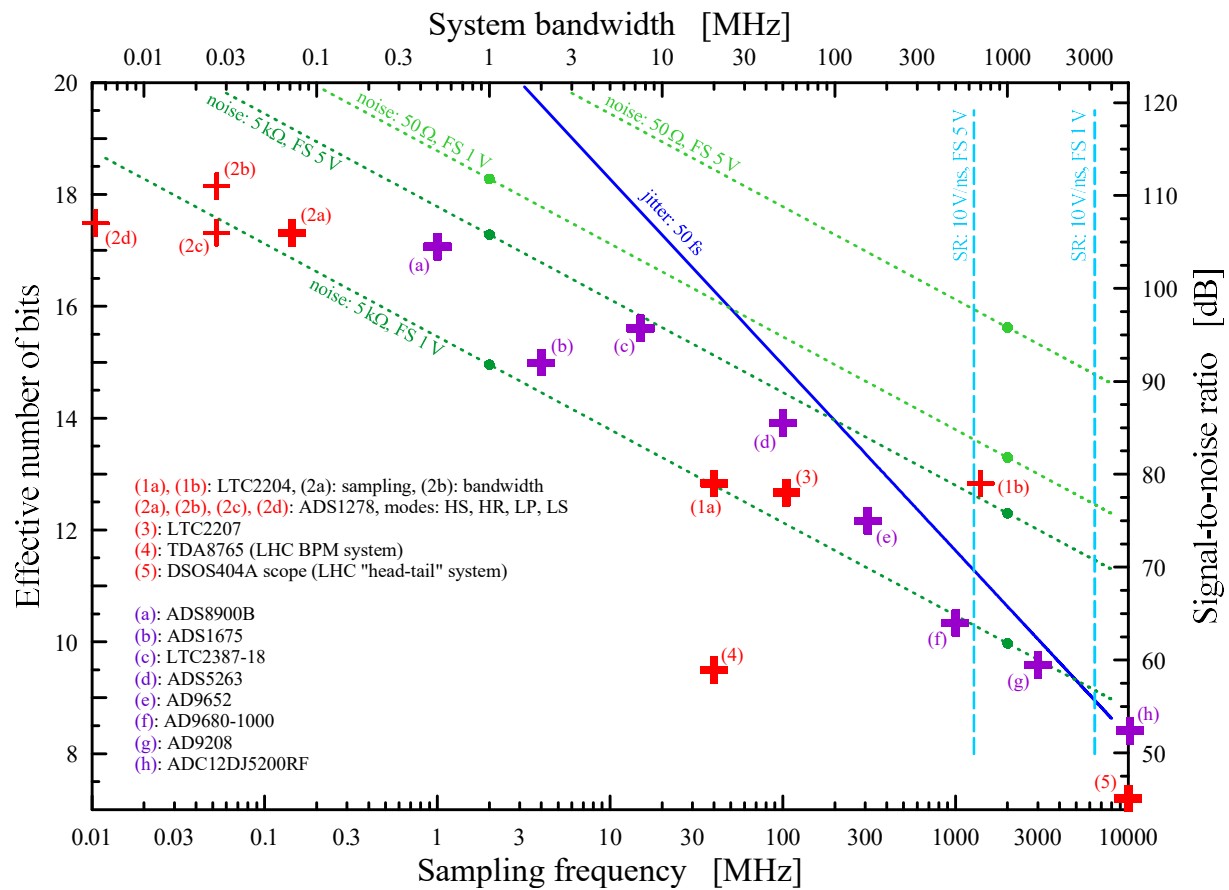

**Fig. 21:** ADC datasheet ENOBs and SNRs along with marked slew-rate, noise and jitter limitations described in this paper. Red crosses correspond to ADCs mentioned in the text while purple ones indicate selected ADCs with an outstanding performance in 2019.

Slew rate limitations discussed in chapter 3.1 are shown on the plot with two vertical light blue dashed lines indicating the bandwidths corresponding to the slew rate of 10 V/ns and two full-scale amplitudes, 10 V and 1 V.

The signal-to-noise limitations described in chapter 3.3 are illustrated on the plot with green dotted lines. The lines correspond to the noise resistances and amplitudes listed in Tables 1 and 2, with their values marked on the lines with dots for the analog bandwidths of 1 MHz and 1 GHz. The lines indicate the SNR change of 10 dB per bandwidth decade and most of the ADCs on the plot are located between or close to the two lines corresponding to the combination of a 5 kΩ noise resistor with 5 V and 1 V full-scale amplitudes. It can be seen that only two ADCs have SNRs much below the "5 kΩ, FS 5 V" line. The first one is TDA8765 used for the LHC BPM system, and which was introduced on the marked some 20 years ago. The poorer noise performance of this ADC shows the progress that has been made in ADC technology since. The second outlying case corresponds to the oscilloscope of the LHC "head-tail" system. This point is quite special, as it concerns not only the ADC, but the oscilloscope as a complete system, including the performance of the input stage with attenuators and amplifiers that introduce their own noise.

The jitter limitation discussed in chapter 3.4 is marked on the plot with blue solid line corresponding to 50 fs jitter. The line function was evaluated by rearranging (27) to calculate the number of bits as a function of frequency, with the first-order slope of 3.3 bits or 20 dB per decade. It can be seen that only one ADC is well above the line, corresponding to the aforementioned special location on the plot of the LTC2204 when marked according to its analog bandwidth. Located slightly above the line is only the fastest and most modern ADC12DJ5200RF, with its first datasheet released four months prior to writing this paper, which is again a sign of constant progress in the ADC performance.

Please note that the overview plot of Fig. 21 indicates only the noise performance of the ADCs expressed either in the effective number of bits or the signal-to-noise ratio and the plot was prepared with a number of assumptions and simplifications, most of which were explicitly mentioned. ADC noise performance is often very important in beam instrumentation systems but not always the most important factor. In order to illustrate one such a case, let's consider two ADCs from the plot with similar sampling rates around 100 MHz, namely LTC2207 (3) and ADS5263 (d). The first one is an older chip with SNR of 78 dB, so if it limits the performance of a beam instrumentation system, then one may consider replacing it by the newer ADS5263 with SNR of 85.5 dB, giving a potential SNR improvement of 7.5 dB, that is a factor of 2.4. For some cases it could be an important improvement, providing that the factor limiting the system performance is indeed the signal-to-noise ratio of the ADC. However, if the LTC2207 is used in a tune measurement system employing FFT spectra with many points, then its replacement by the ADS5263 would be a major downgrade, as shown in Fig. 22, which compares the datasheet spectra from both ADCs. Unfortunately, the spectra are not of equal length, but as explained in chapter 2, this fact changes only the spectra noise floor, while in this case the most important parameter is the level of spurious spectral components, quantified in ADC datasheets with spurious-free dynamic range, mentioned in chapter 2. As the SFDR of the LTC2207 is 23 dB higher (a factor of 14) than the SFDR of ADS5263, the LTC2207 is by far better choice for beam instrumentation systems employing spectral analysis.

Power concerns discussed in chapter 3.2 are illustrated in Fig. 23, presenting a plot with powers dissipated by the selected ADCs versus the sampling frequency in the same six-decade range as the overview plot in Fig. 21. The power scale spans over more than three orders of magnitude, from 5 mW to 10 W.

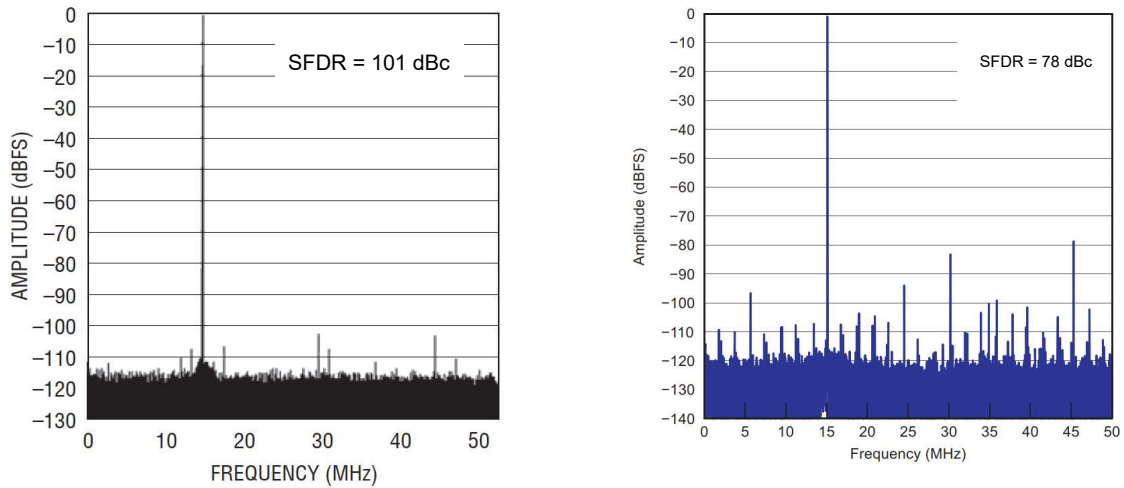

**Fig. 22:** Datasheet spectra of LTC2207 (left) and ADS5263 (right). Measurement conditions for LTC2207 (ADS5263): FFT length of 64 K (32 K) points, 105 MHz (100 MHz) sampling, input signal level of −1 dBFS (−1 dBFS), input signal frequency of 14.8 MHz (15 MHz).

A few ADCs presented on the plot have more than one channel and for such cases the total dissipated power was divided by the number of ADC channels. A special example is ADC12DJ5200RF, which appears on the plot twice. This ADC can operate as a single-channel converter with the fastest sampling of 10.4 GHz (h1) or a dual-channel ADC with the maximal sampling of 5.2 GHz (h2). Such a specification suggests how the faster sampling is achieved: the sampling of the two ADCs is interleaved to trade the number of channels for the sampling speed. Please note that this trick can be used only if the sampling of the individual ADCs is accurate enough to fulfil the jitter requirements of the faster interleaved sampling. The interleaved-sampling technique has been employed by the oscilloscope industry already for some years and nowadays very fast sampling rates are achieved by expanding this technique for many ADCs.

The distribution of symbols in Fig. 23 was fitted with a function

$$\text{dissipated power [W]} = 0.09 \cdot (\text{sampling frequnecy [MHz]})^{0.43} \qquad (33)$$

shown in the plot as a solid green line. The fit is quite close to a simple function plotted for reference as an orange dashed line

$$\text{dissipated power [W]} = 0.1 \sqrt{\text{sampling frequnecy [MHz]}} \qquad (34)$$

The fitting function (33) is completely empirical and based on only 16 data points. However, it is interesting to see that the fit is fairly close to the simple function (34), giving one order magnitude power change for two order magnitude variation of the sampling frequency.

Please note that all of the ADCs shown in the plot with sampling rates beyond 1 GHz have power dissipations beyond 1 W and the fastest one dissipates almost 6 W. Such powers should be addressed at an early stage of the PCB design or during the planning of the accommodation of commercial modules. Even more attention should be payed if the ADC boards have small dimensions, resulting in large power dissipation densities. A particular challenge may be encountered if ADCs with such large dissipations are planned to be used for precision measurements, requiring a particular symmetry between the system channels or demanding long-term stability of the ADC gains and offsets.

As discussed in this paper, with the excellent ADCs available on the market at reasonable prices, simple architectures of beam instrumentation systems should probably be considered first, with the ADCs placed in the signal processing chain close to the beam sensor. With adequate filtering, amplification or attenuation of the sensor signals such a simple approach may allow finding ADCs that satisfy all requirements of at least some beam instrumentation systems. Such approaches minimising analog processing and favouring digital treatment of ADC data should be possible for faster and faster signals with the progress of ADC technology.

On the other hand, requirements for beam instrumentation systems also become more and more challenging, at least for some machines. This is why simplistic architectures of beam instrumentation systems will not be always possible. Even the best ADCs may not satisfy the requirements for the speed and resolution at the same time, the system price or the acceptable throughput of the generated data. In such cases more complex analog signal processing will still be necessary. A few examples of such processing are given in this paper. As explained, an adequate analog processing scheme may decrease the signal bandwidth before the ADC, even by orders of magnitude, which in turn enables the option of using slower ADCs with higher resolution and generating slower data streams. In general, as discussed in the paper, reducing the sampling speed is beneficial for almost all aspect of beam instrumentation, except the time resolution itself. Lower sampling rates help to achieve smaller noise, lower power dissipation and relaxed requirements for the clock jitter and slew rates. Also designing boards with slower ADCs, even of higher resolution, is by far simpler and faster than designing boards with very fast ADCs. All these factors contribute to smaller cost of beam instrumentation systems with slower ADCs.

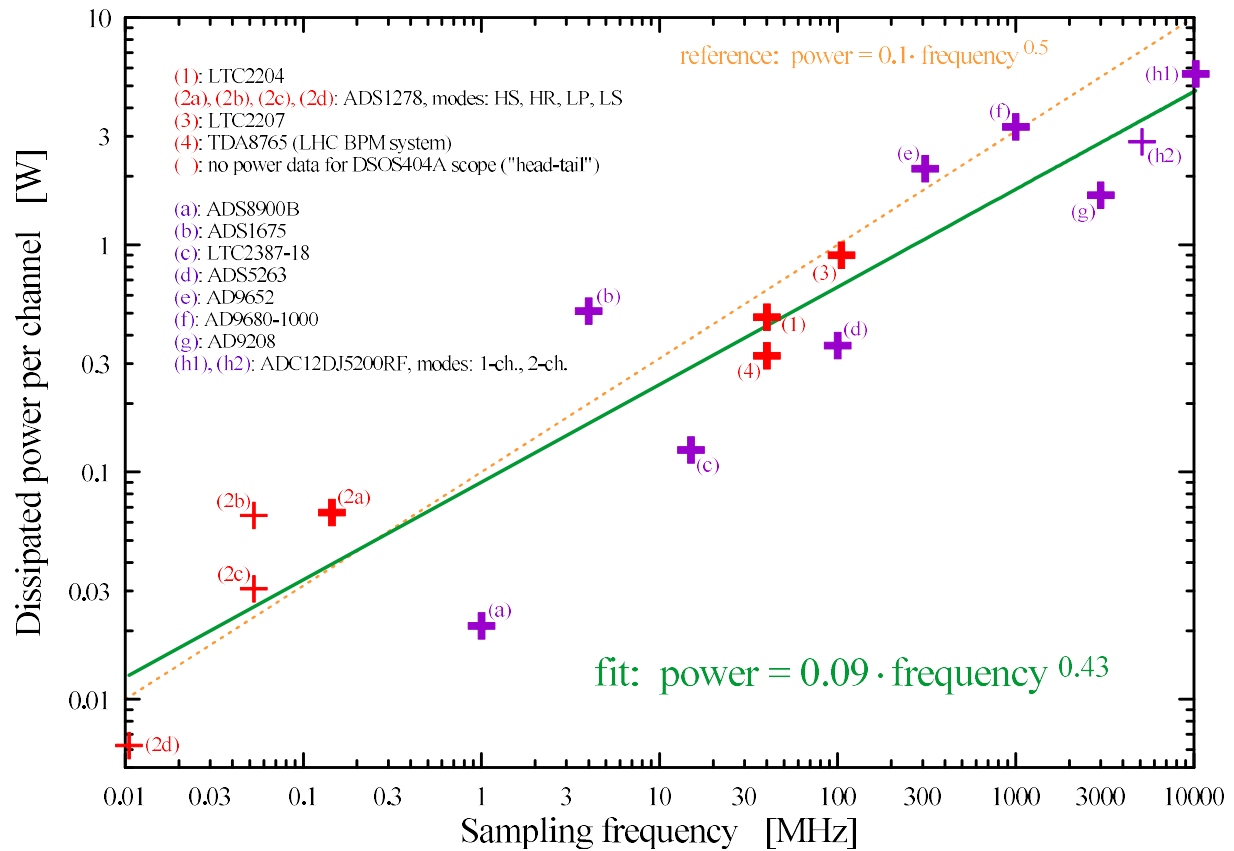

**Fig. 23:** Power dissipation versus the sampling frequency for selected ADCs, along with a fit function (green solid line) and a simple reference function (orange dashed line).

In the case of more challenging beam instrumentation systems even reducing the signal bandwidth before the ADC to the strict minimum by clever analog processing may not allow us to find an ADC with an acceptable combination of sampling speed, resolution and cost. In such cases one can divide the system total dynamic range into two or more sub-ranges. Advantages and consequences of such alternatives are also discussed in the paper.

Practical considerations for choosing an ADC chip and options for obtaining ADC boards are deliberated in the last chapters of this paper. It is hoped that matters discussed there will be of some help in designing beam instrumentation systems.

## Acknowledgment

The final version of this paper was prepared with the valuable help of Tom Levens and Michał Krupa from the CERN Beam Instrumentation Group.

## Literature

[1] W. Kester (Editor), *The Data Conversion Handbook, 2005*, Analog Devices, https://www.analog.com/en/education/education-library/data-conversion-handbook.html

[2] P. Horowitz, W. Hill, *Art of Electronics*, 3rd edition, Cambridge University Press, 2015

[3] U. Tietze, Ch. Schenk, *Electronic Circuits: Handbook for Design and Application*, 2nd edition, Springer, 2008.

[4] J. Belleman, *From analog to digital*, Proceedings of Beam Diagnostics CAS 2008, Dourdan, France, pp. 281 – 316, http://cdsweb.cern.ch/record/1071486/files/cern-2009-005.pdf

[5] J. Belleman, *From analog to digital*, Lecture on Beam Diagnostics CAS 2008, Dourdan, France, http://cas.web.cern.ch/sites/cas.web.cern.ch/files/lectures/dourdan-2008/belleman.pdf

[6] W. Kester, *Understand SINAD, ENOB, SNR, THD, THD + N, and SFDR so You Don't Get Lost in the Noise Floor*, MT-003 Tutorial, Analog Devices, https://www.analog.com/media/en/training-seminars/tutorials/MT-003.pdf